\documentclass[fleqn,usenatbib]{mnras}
\usepackage{newtxtext,newtxmath}
\usepackage[T1]{fontenc}
\usepackage{graphicx}	
\usepackage{amsmath}	

\title[Stochastic process for pulsar timing noise]{
    Stochastic processes for pulsar timing noise: fluctuations in the internal and external torques
}

\author[M. Antonelli et al. ]{
    Marco~Antonelli$^{1}$
    \href{https://orcid.org/0000-0002-5470-4308}{\includegraphics[scale=.5]{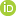}}, 
    Avishek~Basu$^2$
    \href{https://orcid.org/0000-0002-4142-7831}{\includegraphics[scale=.5]{ORCID-iD_icon-16x16.png}},
    Brynmor~Haskell$^{3}$
    \href{https://orcid.org/0000-0002-8255-3519}{\includegraphics[scale=.5]{ORCID-iD_icon-16x16.png}}
    \\
    $^1$CNRS/IN2P3, Laboratoire de Physique Corpusculaire, 14050 Caen, France
    \\
    $^2$Jodrell Bank Centre for Astrophysics, School of Physics and Astronomy, University of Manchester, M13 9PL, Manchester, United Kingdom 
    \\
    $^3$Nicolaus Copernicus Astronomical Center, Polish Academy of Sciences, ul. Bartycka 18, Warszawa, Poland 
    \\
    E-mail: $\,$\url{antonelli@lpccaen.in2p3.fr} $\,$(M.$\,$A.)$\,$, \url{avishek.basu@manchester.ac.uk} (A.B.), \url{bhaskell@camk.edu.pl} (B.H.)
    }

\begin{document}
\date{}
\maketitle

\begin{abstract}
Young pulsars deviate from a perfectly regular spin-down by two non-deterministic phenomena: impulsive glitches and timing noise. Both phenomena are interesting per se, and may provide insights into the superfluid properties of neutron stars, but they also act as a barrier to high-precision pulsar timing and gravitational wave experiments. We study a minimal stochastic model to describe the spin-down of a multicomponent neutron star, with fluctuations in both the internal and external torques. The power spectral density and timing noise strength of this kind of model can be obtained analytically, and compared with known results from pulsar timing observational campaigns. In particular, the presence of flat regions of the power spectral density can be interpreted as a signature of the presence of internal superfluid components. We also derive the expected scaling of the timing noise strength with the pulsar's rotational parameters (or characteristic age). Therefore, the present framework offers a theoretical guideline to interpret the observed features of timing noise in both single pulsars and across the pulsar population.
\vspace{1.5cm}
\end{abstract}




\section{Introduction}

The continuous monitoring of rotation powered radio pulsars in a timing campaign reveals both sudden glitches (impulsive spin-up events), as well as a long-timescale smooth wandering of their rotation period known as timing noise \citep{alessandro_review_1996}. 
Timing noise is a type of rotational irregularity observed in several pulsars, where the measured pulse time of arrival (TOA) wander stochastically about the expected TOA obtained from best-fit timing model via fitted ephemeris \citep{boynton_apj_1972,corders_JPLpulsar_III_1985}. 
It is often characterised as a random walk in the rotational phase, angular velocity, or torque \citep{Groth_1975ApJ_III}; its Fourier spectrum is always red, implying a process autocorrelated on a time-scale of a few months to years \citep{Cordes1980_III,vanHaasteren2013,Liu_2019MNRAS}, 
 {making it a distinct phenomenon from the so-called ``jitter'' noise caused by variations in pulse shape at the single-pulse level \citep{Taylor1975ApJ,Rickett1975ApJ,Lam_NANOGrav_2019}.}

Timing noise is likely to be a genuine feature of the rotational dynamics of a pulsar \citep{hobbs_2010MNRAS}.
Hence, this phenomenon has been attributed to various possible mechanisms at work in the neutron star interior or within its surroundings like unresolved microglitches \citep{corders_JPLpulsar_III_1985,Janssen_slow30_2006A}, recovery from unseen glitches \citep{alpar_noise_1986,Johnston1999MNRAS}, changes in the magnetosphere \citep{Lyne+2010,Shaw+2022}, variable coupling between the crust and liquid interior \citep{alpar_noise_1986,jones_noise_1990MNRAS}, turbulence \citep{MelatosLink2014} and stochastic variations in the star's shape and fluctuations in the spin-down torque, see \citep{alessandro_review_1996} and references therein. 

In this paper we set a up a framework to study how fluctuations in both the \emph{internal} and \emph{external} torques in a rotating neutron star can affect the spin wandering of the observed phase. To do this, we rely on the natural extension of the seminal model proposed by \citet{Baym+1969} for the rotational dynamics of  a two-component neutron star (see \citealt{montoli2020A&A} for a discussion of its three-component extension): we complement it by introducing additive noise in the braking torque and in the internal friction between the components. 

The minimal model of Baym and collaborators proved to be an invaluable tool to interpret glitches, to the point that both linear \citep{Pizzochero2020A&A,Sourie_vela_2020MNRAS,montoli2020A&A} and nonlinear \citep{guerci2017MNRAS,celora2020MNRAS} extensions have been developed and widely used to study the spin-up of a glitch and its subsequent recovery, including the general relativistic corrections \citep{sourie2017MNRAS,antonelli2018MNRAS,gavassino2020MNRAS}. In particular, the signature of an overshoot in the phase residuals recorded during spin-up phase of a glitch in the Vela pulsar \citep{ashton2019Nat,montoli2020A&A} forces us to extend the original two-component model of \citet{Baym+1969} to account for more degrees of freedom in the interior of a pulsar. This can be done in two ways, by considering that the superfluid can develop continuous differential rotation \citep{haskell_pizzochero_2012,antonelli2017,Graber2018ApJ}, or by adding a third rigid component \citep{Pizzochero2020A&A}. Given this observational signature, we will consider a generic number of internal components. 

This kind of stochastic modelling can serve as a theoretical tool to study and derive the expected features of timing noise, with particular attention to its spectral properties (like the presence of corner frequencies that indicate a spectral turnover), in the hypothesis that the origin of noise is intrinsic to the pulsar. 
In fact, whatever the origin of noise, the presence of loosely coupled superfluid layers may leave its imprint on the spectral properties of the detected noise. 
Hence, we analytically derive the scaling of the timing noise strength parameter (the  {root mean square} of the time residuals, see e.g. \citealt{Shannon_noiseMSP_2010}) with the rotational parameters of the pulsar and the physical input of the stochastic model, like the moments of inertia and the superfluid mutual friction parameters.

A similar stochastic approach has also been recently studied by \citet{meyers2021a,meyers2021b} in the context of the possible detection of continuous gravitational wave emission. The main difference with our work is that they consider two external torques acting on two components, while our theoretical setting is for a generic number $m+1$ of internal components driven by one external torque and $m$ internal ones (further extension of the framework is given in App.~\ref{app1}).
Clear distinction between the nature of the torques makes the algebraic results more transparent and allows us to select only the physical results for the timing noise spectrum. 

The paper is organised as follows. In Sec. \ref{theory} we describe the general construction of the minimal stochastic extension of the seminal rigid model of \citet{Baym+1969}, for $m$ superfluid components (the special cases $m=1,2$ are solved explicitly).
In Sec. \ref{strength_measure} we discuss a prescription to set the strength of the stochastic fluctuations in the internal and external torques. Sec. \ref{sec:turnover} is devoted to deriving the spectral properties of the timing noise expected for the models with $m=1,2$.  
We show how to extract the timing noise strength $\sigma$ of \citet{Shannon_noiseMSP_2010} from simulations in Sec. \ref{sec:tnstr}: the scaling of the noise timing noise strength with the rotational parameters of a pulsar (see, e.g. \citealt{Parthasarathy2020MNRAS_I}) is also discussed. 
Finally, in Sec. \ref{sec:num}, we perform numerical simulations with the double intent of exploring the role of the various model parameters and validating the correctness of our analytical results. 
{  The specific two-component model of \citet{meyers2021a} is discussed within our formalism in App.~\ref{app1}. App.~\ref{appOU} is devoted to the interesting limit in which our model reduces to an Ornstein–Uhlenbeck process.}  
\\
\\
{\emph{Notation:} Throughout the paper, the instantaneous angular velocity will be indicated as $\Omega_p(t)$. We adopt the convention that $\Omega>0$ and $\dot\Omega>0$ are constant values for the \emph{measured} angular velocity and the absolute value of the secular spin-down rate of a given pulsar: $\Omega \approx \Omega_p(t)$ and $\dot\Omega \approx |\dot\Omega_p(t)|$. The parameter $|\dot{\Omega}_{\infty}|>0$ always indicates the strength of the spin-down torque normalised over the pulsar's total angular momentum. Practically speaking, $\dot\Omega$ is an estimate - obtained from long pulsar timing observations - of the unknown torque strength $|\dot{\Omega}_{\infty}|$. Hence, to keep notation light, we use $\dot\Omega$ in place of $|\dot{\Omega}_{\infty}|$ in expressions where the exact value of the spin-down torque is unimportant. } 

\section{Multi-component systems with additive noise}
\label{theory}


Building on the seminal work of \citet{Baym+1969}, we consider a minimal model for the evolution of $m+1$ rigid internal components with angular velocities 
\begin{equation}
 \mathbf{\Omega}_t = \left( \Omega_p(t), \Omega_1(t),...,\Omega_m(t) \right)  \, . 
\end{equation}
We identify its first component $\Omega_p(t)$ with the observable angular velocity, which is typically identified with the rotation of normal matter in the star and the magnetosphere \citep{Heintzmann_AA_1973,easson1979}, where the signal originates. 
The other $m$ variables refer to different internal superfluid components (e.g., $m=1$ for the original model of Baym and collaborators, $m=2$ for the Vela's spin-up model described in \citealt{Pizzochero2020A&A}): 
{ {in principle, they may be different superfluid species coexisting in the same region, as well as the same kind of superfluid extending in $m$ spatially non-overlapping layers
\footnote{
   { { The general setting presented here is agnostic about the actual disposition of the superfluid components inside the star, the only requirement being the constraints \eqref{sumx} and \eqref{Bnull}. However, in Sec. \ref{m2} we explicitly consider the particular model of \citet{montoli2020A&A}, where there are two distinct superfluid regions.}}
}.}}

Each of the $m+1$ components contributes to a certain fraction $\mathbf{x}=(x_p, x_1,...,x_m)$ of the total moment of inertia. We do not consider the effect of possible starquakes, implying that both the total moment of inertia and the moment of inertia ratios $\mathbf{x}$ are constant. Therefore, only $m$ out of the $m+1$ values of $\mathbf{x}$ are independent, 
\begin{equation}
\label{sumx}
    \sum_{i=1...m} x_i =1-x_p\, .
\end{equation} 
The ratio $x_p$ includes all components, including some of the superfluid components (if any), which are strongly coupled with the normal matter (non-superfluid component) of the star. 

Regardless of the nature of the internal torques, the total angular momentum\footnote{
    We will use the terms ``angular momentum'' and ``torque'' even if they are divided by the total moment of inertia of the star.
} 
$L(t) = \mathbf{x}^\top\cdot \mathbf{\Omega}$ is affected only by the continuous action of the external torque, 
\begin{equation}
\label{Ldot}
  \dot{L} \, = \, \mathbf{x}^\top \cdot \dot{\mathbf{\Omega}} \, = 
  \, \mathcal{T}_\infty + \eta_t^{\infty}    \, ,
\end{equation}
where $\mathcal{T}_\infty$ is the deterministic part of the total braking torque (arising from electromagnetic and, possibly, gravitational wave emission), while $ \eta_t^{\infty} $ is its fluctuating part. 
Given the above setting, we model the pulsar's evolution in terms of a  It\^{o} equation of the kind 
\begin{equation}
\label{ito}
    d\mathbf{\Omega}_t = (B \, \mathbf{\Omega}_t + \mathbf{A}_t)dt 
    + M d{\mathbf{W}_t} \, , 
\end{equation}
that is the natural stochastic generalisation of the multicomponent models for glitches with linear coupling between the components \citep{Baym+1969,montoli2020A&A}. 
Here, $B$ and $M$ are $(m+1)\times (m+1)$ real matrices and $\mathbf{A}_t$ is a (possibly time dependent) real vector describing the braking torque.  
The term $d{\mathbf{W}_t}$ is the increment of a standard $(m+1)$-dimensional Wiener process, where all the components of $\mathbf{W}_t$ are independent identically distributed (i.i.d.) normal random variables of zero mean and variance $t$, which allows us to model the fluctuating part of the external torque in \eqref{Ldot}, as well as possible fluctuations of the other $m$ internal torques (for a total of $m+1$ independent noise sources). In fact, the mutual transfer of angular momentum between each of the $m$ superfluid components and the observable one may undergo stochastic fluctuations because of the intrinsic noisiness in the vortex-mediated processes \citep{alpar_noise_1986} or internal turbulence \citep{MelatosLink2014}.

Equation \eqref{ito} is very general, but there are physical constraints it should satisfy: the first is that it must be consistent with \eqref{Ldot}, the second is that there should be no internal friction between components that have the same velocity. Therefore, the vector $\mathbf{\Omega}_t=(\Omega \, ,...,\Omega)$ must be a right null eigenvector of $B$,
\begin{equation}
\label{Bnull}
    B (\Omega \, ,...,\Omega) =0 \, ,
\end{equation}
implying that $B$ is not invertible (at corotation, there can not be any deterministic transfer of angular momentum between the components due to friction). 

The It\^{o} process in \eqref{ito} can also be written in the equivalent Langevin form,
\begin{equation}
\label{langevin}
\begin{split}
    &\dot{\mathbf{\Omega}}_t = B \, \mathbf{\Omega}_t + \mathbf{A}_t + M \dot{\mathbf{W}}_t  \, , 
    \\
    & \langle \dot{\mathbf{W}}_t \rangle  = 0 
    \\
    & \langle \dot{W}^i_t \dot{W}^j_s\rangle  = \delta_{ij} \delta(t-s) 
\end{split}
\end{equation}
where the white noise $\dot{\mathbf{W}}_t$ is the formal time derivative of the standard Wiener process, $d\mathbf{W}_t=\dot{\mathbf{W}}_t\,dt$, see e.g. \citet{Gardiner1994book}. 
Clearly, the system in \eqref{langevin} must be consistent with the fundamental requirement \eqref{Ldot}, so that $\mathbf{A}_t$, $B$ and $M$ can not be set independently, but they should satisfy the constraint
\begin{equation}
\label{balena}
      \mathbf{x}^\top \cdot (B \, \mathbf{\Omega}_t + \mathbf{A}_t + M \dot{\mathbf{W}}_t) =  
       \mathcal{T}_\infty + \eta_t^{\infty} \, .
\end{equation}
The above equation defines a physical property of the coupling matrix $B$: internal torques should satisfy the action-reaction law. In fact, to guarantee that \eqref{balena} is consistent with \eqref{Ldot}, $\mathbf{x}$ must be a null left eigenvector of $B$, 
\begin{equation}
\label{actionreaction}
  B^\top \mathbf{x}=0   \qquad \quad \text{(action-reaction property)} \, ,
\end{equation}
and, on the other hand, the total external torque must be given by
\begin{equation}
\label{quisquilie}
    \mathbf{x}^\top \cdot \mathbf{A}_t =  \mathcal{T}_\infty 
    \qquad \qquad
    \mathbf{x}^\top M\dot{\mathbf{W}}_t = \eta_t^{\infty} \, .
\end{equation}
The solution to \eqref{ito} is not formally different from the usual solution for an ordinary differential equation of the same form, 
\begin{equation}
\label{ito_sol}
     \mathbf{\Omega}_t = e^{B t}\mathbf{\Omega}_0 +\int_0^t \!\!e^{B (t-z)} \mathbf{A}_z \, dz +\int_0^t \!\!e^{B (t-z)} M \, d\mathbf{W}_z \, .
\end{equation}
For each $t$, the stochastic integral above defines a random variable of zero average, so that the expected value $\langle \mathbf{\Omega}_t \rangle$ coincides with the deterministic drift of $\mathbf{\Omega}_t$, 
\begin{equation}
\label{sol_det}
     \langle \mathbf{\Omega}_t \rangle = e^{B t}\langle \mathbf{\Omega}_0 \rangle +
     \int_0^t \!\! e^{B (t-z)} \mathbf{A}_z \, dz\, .
\end{equation}
Therefore, by imposing a deterministic initial condition $\mathbf{\Omega}_0=\langle \mathbf{\Omega}_0 \rangle$, we can define the angular velocity residual as 
\begin{equation}
\label{ang_vel_res}
     \delta\mathbf{\Omega}_t = \mathbf{\Omega}_t  -  \langle \mathbf{\Omega}_t \rangle =
    \int_0^t \!\! e^{B (t-z)} M \, d\mathbf{W}_z \, ,
\end{equation}
that is the solution of the Langevin equation
\begin{equation}
\label{langevin2}
\delta\dot{\mathbf{\Omega}}_t =  B \, \delta\mathbf{\Omega}_t +M\dot{\mathbf{W}}_t \, .
\end{equation}
We are interested in the statistical properties of the residuals $\delta\mathbf{\Omega}_t$, in particular of its first observable component $\delta \Omega^p_t$. 
Given the solution \eqref{ang_vel_res}, the average and the correlation matrix read\footnote{
    Since we imposed a deterministic initial condition, we have that $ \delta\mathbf{\Omega}_0=0$. If this hypothesis is dropped, then the results in \eqref{isometry} gain an extra term.
}
\begin{equation}
\label{isometry}
    \begin{split}
    &\langle  \delta{\Omega}^i_t  \rangle =  0 \qquad \forall \, i = p,1, ..., m 
    \\
    &\langle \delta\mathbf{\Omega}_t \delta\mathbf{\Omega}_u^\top  \rangle 
     = \int_0^{\min(t,u)} \! \! dz \,\, e^{B(t-z)}M M^\top e^{B^\top(u-z)}
     \, .
    \end{split}
\end{equation}
Since the correlation matrix is a function of both times $t$ and $u$, the process is not wide-sense stationary: for  $u<t$ and $t=u+\Delta t$, the matrix $\langle \delta\mathbf{\Omega}_{u+\Delta t} \delta\mathbf{\Omega}_u^\top  \rangle $ depends on $u$. Therefore, to calculate the Power Spectral Density (PSD) of the process we can \emph{not} rely on the Wiener–Khinchin theorem. However, it is still possible to perform the Fourier transform on the Langevin equation and use the frequency representation of the white-noise \citep{rice1944}. 
First, we introduce the Fourier representation 
\begin{equation}
\label{fourier}
 \delta \mathbf{\Omega}_t = \int_{-\infty}^{\infty} \!e^{i \omega t } \delta\mathbf{\Omega}_\omega \, \dfrac{d\omega}{2 \pi}   
\end{equation}
and define the matrix  
\begin{equation}
\label{psd}
 P(\omega,\omega') =  \langle \delta\mathbf{\Omega}_{\omega} \, \delta\mathbf{\Omega}^\dagger_{\omega'} \rangle   \, ,
\end{equation}
 where $\delta\mathbf{\Omega}^\dagger_\omega$ is the transpose complex conjugate of the vector $\delta\mathbf{\Omega}_\omega$. In particular, we are interested in the $i=j=p$ component for $\omega = \omega'$, that is the PSD of the signal associated to the observable component:
\begin{equation}
\label{P00def}
 P_{p}(\omega) \propto  \langle \, | \delta \Omega_p (\omega) |^2 \, \rangle   \, .
\end{equation}
From \eqref{langevin2} we have 
\begin{equation}
    \delta\mathbf{\Omega}_\omega = (i \omega \, \mathbb{I}- B)^{-1} M\,\dot{\mathbf{W}}_\omega \, ,
\end{equation}
that, together with the basic properties of the Wiener process,
\begin{equation}
\label{rice}
    \begin{split}
    &\langle \dot{ W }^i_\omega  \rangle = 
    \int_{-\infty}^{\infty}\! dt \, e^{-i\omega t}\langle \dot{ W }^i_t  \rangle = 0 \qquad \forall \, i = p,1, ..., m  
    \\
    &\langle \dot{\mathbf{W}}_\omega \dot{\mathbf{W}}_{\omega'}^\dagger \rangle 
    = 2 \pi \, \delta(\omega-\omega')\,\mathbb{I}
     \, ,
    \end{split}
\end{equation}
can be used to calculate the matrix in  \eqref{psd}:
\begin{equation}
\label{riceeee}
     P(\omega,\omega') = 2 \pi \,(i \omega\,\mathbb{I} - B)^{-1} M M^\top 
     (-i \omega' \,\mathbb{I}- B^\top)^{-1} \delta(\omega'-\omega) \, . 
\end{equation}
Therefore, $P_{p}(\omega)$ will contain a divergent Dirac delta factor that is, however, just a consequence of introducing the (formal) Fourier representation \eqref{fourier} for a non-normalizable signal \citep{priestley1965}: following \cite{Gardiner1994book}, we define the PSD for the observable component as 
\begin{equation}
\label{rice00}
     P_{p}(\omega) = \left[ (i \omega\,\mathbb{I} - B )^{-1} M M^\top 
     (-i \omega \,\mathbb{I}- B^\top)^{-1} \right]_{pp}   \, , 
\end{equation}
where the subscript $pp$ means that only the $i=j=p$ component of the matrix defined within the square brackets is taken. Note that $P_{p}(\omega)$ has the physical dimensions expected for a PSD, namely that of a power (i.e., the squared amplitude of the signal, $|\delta\Omega^p_t|^2$) per unit angular frequency $\omega$. Since both $\delta\Omega^p_t$ and $\omega$ have the dimensions of rad/s, also the PSD in \eqref{rice00} is measured in rad/s (a concrete example will be given in Fig. \ref{fig_psd_vel}).

\subsection{Evolution of the total angular momentum and rigid rotor~$(m=0)$  }

Before moving to more complex cases, let us briefly discuss the evolution of the total angular momentum residue $\delta L_t$ for $m>0$. Its evolution is given by
\begin{equation}
\label{lconst}
\delta L_t = \mathbf{x}^\top \cdot \delta\mathbf{\Omega}_t 
\qquad \quad \delta\dot{L}_t = \eta_\infty \, ,
\end{equation}
meaning that it is unaffected by the possible presence of internal torques, as it should be. 
Its associated power spectral density  $P_{L}(\omega) = \langle |\delta L_\omega|^2 \rangle$ can be found by considering that 
\begin{equation}
    \delta L_\omega = \mathbf{x}^\top \cdot \delta\mathbf{\Omega}_\omega 
    \qquad \qquad 
    i\omega \, \delta L_\omega = \eta_\omega = \mathbf{x}^\top M\dot{\mathbf{W}}_\omega
    \, .
\end{equation}
Therefore, we recover the usual result for a pure Wiener process\footnote{
    The PSD in \eqref{PSD_LL} can also be obtained from the matrix in \eqref{riceeee} after double contraction with $\mathbf{x}$: thanks to the action-reaction property \eqref{actionreaction} we have that~$(\omega\,\mathbb{I} - B^\top )^{-1} \, \mathbf{x} =\mathbf{x}/\omega$.
    },  
\begin{equation}
    \label{PSD_LL}
    P_{L}(\omega) = \langle |\delta L_\omega|^2\rangle =\mathbf{x}^\top(i \omega )^{-1} M M^\top 
     (-i \omega  )^{-1} \mathbf{x}
     = \frac{\sigma_\infty^2}{\omega^2} \, ,
\end{equation}
where we have defined the positive parameter $\sigma_\infty^2$  as 
\begin{equation}
    \label{sigma_inf}
    \sigma_\infty^2 \, = \mathbf{x}^\top  M M^\top \mathbf{x}    \, .
\end{equation}
In the limit in which the star rotates rigidly (i.e,  $m=0$), we have that $L = \Omega_p$,  $B=0$ and $M=\sigma_\infty$. Therefore, the PSD for the observable component in \eqref{rice00} boils down to the one in \eqref{PSD_LL}: 
\begin{equation}
    \label{PSD_L}
    P_{p}(\omega) = \frac{\sigma_\infty^2}{\omega^2} 
    \qquad \quad(\text{if} \quad m=0).
\end{equation}
Therefore, in the limit of rigid rotator, we recover the usual red-noise PSD of a pure Wiener process.

\subsection{Two-component model $(m=1)$}
\label{m1}

The simplest $m=1$ model reads 
\begin{equation}
\begin{split}
& x_p \dot{\Omega}_p =  - \mathcal{T} + \mathcal{T}_\infty 
\\
&  x_1 \dot{\Omega}_1 = \mathcal{T} \, ,
\end{split}
\label{batuffolino}
\end{equation}
where $\Omega_p$ and $\Omega_1$ are the angular velocities of the observed and superfluid components, $x_p$ and $x_1=1-x_p$ are the moment of inertia ratios of each component, see \eqref{sumx}, and $\mathcal{T}$ is the internal torque. 
In the following, for simplicity we will set $ \mathcal{T}_\infty=-|\dot{\Omega}_{\infty}|$, a constant value that sets the secular spin-down rate. 
Due to the presence of fluctuations, the intrinsic value of the external torque $|\dot{\Omega}_{\infty}|$ may slightly differ from the absolute value of the spin-down rate $\dot\Omega$ extracted from a fit to the observed signal\footnote{
    We recall that $\dot\Omega>0$ is a constant benchmark value for the observed absolute value of the secular spin down rate, $\dot\Omega \approx |\dot{\Omega}_{\infty}|$.
}.

Note that we are assuming that the physical mechanism responsible for the spin down acts directly only on the observable component. This is reasonable as long as this mechanism does not directly cause an outward motion of vortex lines. 
The internal torque $\mathcal{T}$ in \eqref{batuffolino} is phenomenologically modelled in terms of vortex-mediated friction between the components as
\begin{equation}
\mathcal{T} = - \frac{x_p \, x_1 }{\tau} (\Omega_1 -\Omega_p) 
= - x_1 2\Omega \mathcal{B} (\Omega_1 -\Omega_p)
\end{equation}
where  $\Omega \approx \Omega_p \approx \Omega_1$ is a constant benchmark value for the observed angular velocity and $\mathcal{B}$ is a dimensionless coupling constant that sets the intensity of the vortex-mediated mutual friction; the connection of all these body-averaged parameters with the local hydrodynamic parameters for nuclear matter is derived in App.~C of~\citet{montoli2020A&A}. 
The timescale 
\begin{equation}
    \label{relax_time}
    \tau = x_p/( 2 \mathcal{B}  \Omega)
\end{equation}
is the observed \emph{relaxation} timescale, namely the timescale with which the $m=1$ system exponentially relaxes back to the steady state when perturbed away from it. 

The system admits a steady state in which the value of the lag $\Omega_1-\Omega_p$ is constant over time: imposing the steady state condition $\dot{\Omega}_p = \dot{\Omega}_1 = -|\dot{\Omega}_{\infty}|$, we have
\begin{equation}
\label{steady}
\begin{split}
    & \Omega_1 -\Omega_p = \tau |\dot{\Omega}_{\infty}|/x_p = |\dot{\Omega}_{\infty}|/(2 \Omega \mathcal{B} ) 
    \\
    &  |\mathcal{T}| = x_1  \, |\dot{\Omega}_{\infty}|
\end{split}
    \quad\qquad \text{(steady state)}
\end{equation}
We can turn \eqref{batuffolino} into a stochastic equation by introducing an additive white noise term to both the external torque and internal torques: 
\begin{equation}
\label{noiseee}
    -|\dot{\Omega}_{\infty}| \rightarrow  -|\dot{\Omega}_{\infty}|+\eta^\infty_t
    \qquad\quad  
    \mathcal{T} \rightarrow \mathcal{T} + \eta^\mathcal{T}_t
\end{equation}
Now, we can recast \eqref{batuffolino} in the form \eqref{langevin} by setting
\begin{equation}
\label{2_comp_mat}
\mathbf{\Omega}=\begin{pmatrix}
    \dot{\Omega}_p \\
    \dot{\Omega}_1
    \end{pmatrix}
\quad 
B = \begin{pmatrix}
    -x_1/ \tau & x_1/ \tau \\
    x_p/ \tau & -x_p/ \tau
     \end{pmatrix}
\quad 
\mathbf{A} =     
    \begin{pmatrix}
     -|\dot{\Omega}_{\infty}|/x_p
     \\
     0
    \end{pmatrix}
\end{equation}
The matrix $M$ can be found by writing the two independent fluctuations in \eqref{noiseee} in terms of the standard white noise  $\dot{\mathbf{W}}_t=(\dot{W}^\infty_t,\dot{W}^\mathcal{T}_t)$ as
\begin{equation}
\label{noises2}
   \eta^\infty_t = \sigma_\infty \, \dot{W}^\infty_t 
    \qquad\quad  
  \eta^\mathcal{T}_t = \sigma_\mathcal{T} \, \dot{W}^\mathcal{T}_t \, ,
\end{equation}
implying that (only the relative signs of the components of $M$ are important)
\begin{equation}
\label{M2}
M = \begin{pmatrix}
    \sigma_\infty/x_p  &  -\sigma_\mathcal{T}/x_p \\
    0                  &   \sigma_\mathcal{T}/x_1
     \end{pmatrix} 
\end{equation}
Differently from \citet{meyers2021a}, {  see App. \ref{pizzone}}, we do not include the additive noise terms as independent fluctuations in the expressions for $\dot{\Omega}_p$ and $\dot{\Omega}_1$. 
Rather, we ascribe them to different independent physical mechanisms: the external (electromagnetic) and internal (vortex-mediated) torques.
By using \eqref{rice00} the PSD for the observable component is
\begin{equation}
\label{P002}
    P_{p}(\omega) = \dfrac{
    (\sigma_\mathcal{T}^2 + \sigma_\infty^2)\tau^2\omega^2 +x_p^2\sigma_\infty^2
    }{x_p^2 (\tau^2 \omega^4+\omega^2)} \, ,
\end{equation}
that reduces to \eqref{PSD_LL} in the double limit $ \sigma_\mathcal{T}^2 \rightarrow 0$,  $x_p \rightarrow 1$. To compare this result with the one for the $m=2$ model described in the next subsection, we may express the PSD in terms of the convenient coupling parameter 
\begin{equation}
\label{pizzobaldo_b}
 b\, =\,2\, \Omega \, \mathcal{B}  \, = \, x_p/\tau  \, , 
\end{equation}
obtaining ($\mu$ and $\xi$ are two parameters that define the PSD)
\begin{equation}
\label{P002_b}
  \begin{split}
  & P_{p}(\omega) \propto \dfrac{1}{\omega^2} \cdot \dfrac{ \mu^2 +  \omega^2 }{ \xi^2 +   \omega^2 } \\
  & \mu^2 = \dfrac{b^2 \,  \sigma_\infty^2}{ \sigma_\infty^2+  \sigma_\mathcal{T}^2} 
 \qquad  \qquad \xi^2 = \dfrac{b^2}{x_p^2}=\dfrac{1}{\tau^2} \, .
 \end{split}
\end{equation}
The above expressions guarantee that the parameters $\xi$ and $\mu$ are ordered as
 \begin{equation}
     0<\mu < \xi \, ,
 \end{equation}
Moreover, $\mu$ and  $\xi$ are the ``corner frequencies'' that define the two spectral turnovers of the PSD, see Section~\ref{sec:turnover}.

\subsection{Three-component model  $(m=2)$}
\label{m2}

The straightforward $m=2$ extension of the model of Baym and collaborators reads, 
\begin{equation}
\begin{split}
& x_p \dot{\Omega}_p =  - \mathcal{T}_1 - \mathcal{T}_2 - |\dot{\Omega}_{\infty}|
\\
&  x_1 \dot{\Omega}_1 = \mathcal{T}_1 
\\
&  x_2 \dot{\Omega}_2 = \mathcal{T}_2 \, ,
\end{split}
\label{batuffolino3}
\end{equation}
where $x_p=1-x_1-x_2$ and
\begin{equation}
\label{linearMF3}
\mathcal{T}_i = - x_i b_i (\Omega_i -\Omega_p) \qquad i=1,2
\end{equation}
This is exactly the deterministic model in equation (1) of \citet{montoli2020A&A} and its general solution and properties are discussed therein.
The coupling parameters $b_i=2 \Omega \mathcal{B}_i$ set the strength of the deterministic part of the mutual friction, while the meaning of the phenomenological dimensionless friction parameter  $\mathcal{B}_i$ in terms of the microscopic input is the one presented in App. C of \citet{montoli2020A&A}. Note that for $m>1$ the relaxation timescales are complicated functions of $x_1$, $x_2$, $b_1$ and $b_2$, see equation (A.19) therein.
Similarly to \eqref{steady}, the steady state lags are, for $i=1,2$, 
\begin{equation}
\label{steady3}
\begin{split}
    &  \Omega_i-\Omega_p = |\dot{\Omega}_{\infty}|/b_i =  |\dot{\Omega}_{\infty}|/(2 \Omega \mathcal{B}_i)
    \\
    &  |\mathcal{T}_i| = x_i  \, |\dot{\Omega}_{\infty}|
\end{split}
    \quad\qquad \text{(steady state)}
\end{equation}
We promote the deterministic model in \eqref{batuffolino3} to a stochastic one by adding fluctuating terms to the external torque and to the two internal torques. 
Following the same procedure outlined in the previous subsection, we define 
$\mathbf{\Omega}=(\dot{\Omega}_p ,\dot{\Omega}_1, \dot{\Omega}_2)^\top $ and
\begin{equation}
\label{m3B}
B = \begin{pmatrix}
\dfrac{-x_1 b_1 -x_2 b_2}{x_p} & \dfrac{x_1 b_1}{x_p} & \dfrac{x_2 b_2}{x_p} \, \\
b_1  & -b_1 &  0   \\
b_2  &  0   & -b_2 
     \end{pmatrix}
\qquad 
\mathbf{A} =     
    \begin{pmatrix}
     - \dfrac{|\dot{\Omega}_{\infty}|}{x_p} \,
     \\
     0
     \\
     0
    \end{pmatrix}
\end{equation}
The matrix $M$ is given by
\begin{equation}
\label{m3M}
M = \begin{pmatrix}
    \sigma_\infty/x_p  &  -\sigma_{\mathcal{T}_1}/x_p &  -\sigma_{\mathcal{T}_2}/x_p \\
   0                  &   \sigma_{\mathcal{T}_1}/x_1 & 0 \\
    0                 &   0 & \sigma_{\mathcal{T}_2}/x_2
     \end{pmatrix}
\end{equation}
Using \eqref{rice00}, the PSD for the observable component is
\begin{equation}
\label{P003}
\begin{split}
   & P_{p}(\omega)   = \dfrac{N_0 +
    N_2 \omega^2 +
        N_4 \omega^4
          }{  
          D_2 \omega^2 + D_4 \omega^4
          + D_6 \omega^6  }
          \propto
\dfrac{n_0 + n_2 \omega^2 + \omega^4}{ d_2 \omega^2 + d_4 \omega^4 + \omega^6  }
          \\
          & N_0 = b_1^2 b_2^2 \sigma_\infty^2
          \\
          & N_2 = b_1^2(\sigma_{\mathcal{T}_2}^2+\sigma_\infty^2 ) +  
                 b_2^2(\sigma_{\mathcal{T}_1}^2+\sigma_\infty^2 )
          \\
          & N_4 = \sigma_{\mathcal{T}_1}^2+\sigma_{\mathcal{T}_2}^2+\sigma_\infty^2 
          \\
         &  D_2 = b_1^2 b_2^2  
          \\
        &   D_4 = b_1^2 (1-x_2)^2+b_2^2 (1-x_1)^2+ 2 b_1 b_2 x_1 x_2
          \\
         &  D_6 = x_p^2
    \\
    &  n_0 = \dfrac{N_0}{N_4}\qquad  n_2=\dfrac{N_2}{N_4} \qquad d_2 =\dfrac{D_2}{D_6} \qquad d_4 =\dfrac{D_4}{D_6}  \, .
\end{split}
\end{equation}
In the double limit $(x_2 \, , \, \sigma_{\mathcal{T}_2}) \rightarrow 0$, the above result reduces to the PSD of the $m=1$  model in \eqref{P002} with $b_1= x_p/\tau$.
Again, we can introduce four positive corner frequencies, $\mu_{\pm}$ and $\xi_{\pm}$, to obtain a form of the PSD that is analogous to the one in \eqref{P002_b}:
\begin{equation}
\label{P003_b}
  \begin{split}
  & P_{p}(\omega) \propto 
  \dfrac{1}{\omega^2} \cdot \dfrac{ (\mu_-^2 +  \omega^2) (\mu_+^2 +  \omega^2) }
  { (\xi_-^2 +  \omega^2) (\xi_+^2 +  \omega^2) } 
  \\
  & \mu_{\pm}^2 = \dfrac{1}{2} \left[ \, n_2 \pm \sqrt{n_2^2-4n_0} \, \right] 
  \qquad 
  \xi_{\pm}^2 = \dfrac{1}{2} \left[ \, d_4 \pm \sqrt{d_4^2-4d_2} \,  \right] 
   \, .
 \end{split}
\end{equation}
Since $0<n_0<d_2$ and $0<n_2<d_4$, which can be seen from the formulae in \eqref{P003}, we have that the four corner frequencies $\mu_{\pm}$ and $\xi_{\pm}$ are naturally ordered as
\begin{equation}
    0<\mu_{-}\leq  \mu_{+} \qquad  \quad
  0<\xi_{-}\leq  \xi_{+}  \qquad  \quad
  0<\mu_{-} < \xi_{+}  
\end{equation} 
Their values define where a spectral turnover should be expected, see section \ref{sec:turnover}.

\section{Setting the strength of fluctuations}
\label{strength_measure}

In this section, we discuss a parametrization of the phenomenological quantities $\sigma_\infty$ and $\sigma_\mathcal{T}$ in terms of the physical parameters appearing in the deterministic part of the model, that have a more transparent hydrodynamic interpretation \citep{montoli2020A&A}. 

To set the amplitude of the fluctuations in both the external and internal torque in terms of the physical parameters we adopt the following scheme. 
Let us consider a generic ``force'' $F=\delta F + F_0$, where $\delta F$ is a white noise process and $F_0$ is deterministic, and its associated physical timescale $T_F$, in the sense that the effect of $F_0$ is to drive the system's evolution on timescales of order $T_F$. 
We can set the strength of $\delta F $ by requiring that its effect, over a time interval of order $T_F$, is a fraction $0<\alpha<1$ of the one produced by the deterministic part $F_0$, 
\begin{equation}
    \int_0^{T_F} \delta F(t) \, dt \approx  \alpha     \int_0^{T_F}  F_0(t) \, dt \, .
\end{equation}
Strictly speaking, the above equation only serves to convey the general idea and is not formally correct (the left hand side is a random variable, but on the right we have a deterministic value). However, the left hand side is typically realised within a few standard deviations from its statistical expectation, that is zero. Therefore, the size of one of its ``typical'' realisations is given by its standard deviation (this would be the rigorous interpretation of the above equation). In view of this, our prescription to set the strength of the noise is
\begin{equation}
\label{abdul}
    \text{std. dev.} \left[ \int_0^{T_F} \delta F(t) \, dt \right] 
    = \alpha     \int_0^{T_F}  F_0(t) \, dt \, .
\end{equation}
In practice, we are requiring that, over the relevant timescale $T_F$, the impulse caused by fluctuations is (typically) a fraction $\alpha$ of the impulse imparted by $F_0$. 
In this way, the noise strength can be parametrized in terms of $\alpha$, which is a relative measure of fluctuations with respect to the deterministic part of the process.  

In the following, to estimate the standard deviation in \eqref{abdul}, we will use the property that the increments
\begin{equation}
\label{bruttostruzzo}
    W^i_{t+\Delta t}-W^i_t \sim \mathcal{N}\left( 0,\sqrt{\Delta t} \right) \qquad \quad \forall \, i , \, \Delta t>0 \, .
\end{equation}
are normally distributed with standard deviation $\sqrt{\Delta t}$.

\subsection{Noise in the external torque} 

In order to set the strength of the external torque  $\sigma_\infty$, it is natural to consider the process in \eqref{Ldot}, namely
\begin{equation}
\label{processo_cesso}
\dot{L} \, dt = - |\dot{\Omega}_\infty| + \sigma_\infty dW^\infty_{t}
\end{equation}
The integration should be performed over the relevant timescale, namely the pulsar's life timescale\footnote{
    For a ``standard'' braking index of 3, the pulsar's characteristic dipole age would be $T/2$.
    } 
$T = \Omega/\dot{\Omega}$, where $\Omega \approx \Omega_p$ and $ \dot{\Omega}  \approx |\dot{\Omega}_\infty|$ are the observed rotational parameters of the pulsar.
Thanks to the fundamental property in \eqref{bruttostruzzo}, we have
\begin{equation}
\int_0^T\!\! \sigma_\infty dW^\infty_{t} =
\sigma_\infty (W^\infty_{t+T}-W^\infty_{t})
\sim \mathcal{N}\left( \, 0 \, ,\, \sigma_\infty \sqrt{T} \, \right) \, ,
\end{equation}
Now, we require that the standard deviation of this process is a fraction $0<\alpha_\infty<1$ of the observed angular velocity $\Omega$,  
\begin{equation}
\label{infinitoso}
    \sigma_\infty \sqrt{T} \, = \, \alpha_\infty  \, \Omega 
    \qquad \qquad 
    \sigma_\infty^2 \, = \, \alpha_\infty^2 \, \Omega \,  \dot{\Omega}
    \, .
\end{equation}
This choice encodes the fact that random fluctuations in the spin down torque can result in a total change of the present angular velocity $\Omega$, of the order of a fraction $\alpha_\infty$ of it, when their cumulative effect over the pulsar's life is taken into account. 
In other words, we are imposing that the typical effect of fluctuations in the external torque piles up to a fraction of the effect of the deterministic spin-down
\begin{equation}
   \left| \int_0^T \mathcal{T}_\infty \, dt \, \right| = 
   |\dot{\Omega}_\infty| T \approx \Omega \, .
\end{equation}

\subsection{Noise in the internal torque} 

We apply the same reasoning to set the strength of the internal torques. For $m=1$, the relevant timescale is in terms of the coupling parameter $b$ in \eqref{pizzobaldo_b}. We demand that  the ``typical'' effect of the fluctuation in a time interval of order $\sim b^{-1}$
\begin{equation}
  \int_0^{b^{-1}} \!\!\! \eta_\mathcal{T}\, dt 
  = \int_0^{b^{-1}}  \!\!\! \sigma_\mathcal{T}  \, dW^\mathcal{T} 
 \sim \mathcal{N}\left( \,0\, , \, \sigma_\mathcal{T} /\sqrt{b} \, \right)
\end{equation}
 amounts to a fraction $0<\alpha<1$ of the effect of the steady state internal torque \eqref{steady}, that is given by
\begin{equation}
\left|  \int_0^{b^{-1}} \!\!\!  \mathcal{T}  \, dt \, \right| 
 =  \frac{|\mathcal{T} |}{b} \, = \,  \frac{x_1 \, |\dot{\Omega}_\infty|}{b} \,
 \approx \, \frac{x_1 \, \dot{\Omega} }{b}  \, .
\end{equation}
Therefore, we require that, 
 \begin{equation}
  \int_0^{b^{-1}} \!\!\! \eta_\mathcal{T}\, dt 
   \sim \mathcal{N}\left( 0\, , \, \alpha \,  x_1 \dot{\Omega} / b\, \right)\, ,
\end{equation}
 namely 
\begin{equation}
\label{psico}
    \sigma_\mathcal{T} /\sqrt{b} = \alpha   \,  x_1   \dot{\Omega} /b
    \qquad \qquad 
    \sigma_\mathcal{T}^2 \,=\, \dfrac{ \alpha^2 \, x_1^2  \,  \dot{\Omega}^2 }{2 \, \mathcal{B} \,  \Omega} 
    \, .
\end{equation} 
For $m=2$, the parametrization of the fluctuations in the external torque is still given by \eqref{infinitoso}, while the analogue of \eqref{psico} is
\begin{equation}
\label{psico3}
    \sigma_{\mathcal{T}i}^2
    = \frac{\alpha_i^2\, x_i^2\, \dot{\Omega}^2}{2 \, \mathcal{B}_i\, \Omega}
    \qquad \, \, 
    0 < \alpha_i <1
    \qquad  \, \, 
    i=1,2
    \, .
\end{equation} 

\section{Spectral flattening and corner frequencies}
\label{sec:turnover}

We can now rewrite the PSD in terms of our ``natural'' choice of the noise variables $\sigma_\infty$ and $\sigma_\mathcal{T}$ derived the previous section. 
In this way, we can link the presence of a possible spectral turnover and the corresponding corner frequencies to the physical properties (i.e., $x_p$, $\tau$, $T$...) of a given pulsar. 

\subsection{Spectral turnover for $\ensuremath{m=1}$}

The PSD for the angular velocity in \eqref{P002_b} is defined in terms of two frequencies $\mu$ and $\xi$. If $\mu \ll \xi$, we can identify 3 power-law regimes with spectral index $-2$, $0$ and $-2$:
\begin{equation}
\label{P002_regimes}
  P_{p}(\omega) \propto 
 \begin{cases}
  \,\,  \mu^2 /(\xi^2 \, \omega^2)   
  \qquad & \omega \ll \mu 
  \\
  \,\, 1/\xi^2
  \qquad & \mu \ll \omega\ll \xi
  \\
  \,\, 1/\omega^2 \qquad & \xi \ll \omega 
  \end{cases}  
\end{equation}
Since the PSD undergoes a spectral turnover around $\omega \approx \xi, \mu$, those two frequencies play the role of corner frequencies, see e.g. \cite{Coles_hobbs_2011MNRAS}.
By looking at \eqref{P002_b}, one may also expect a possible regime~$P_{p}\propto 1/\omega^4 $.
However, given our setting, this possibility is forbidden by the parametrization of the PSD in terms of the physical parameters of the spin-down model: to obtain a region where $ P_{p}\propto 1/\omega^4 $, we should have $\xi \ll \omega \ll \mu $, that is inconsistent with the physical ordering $\mu<\xi$. 
This is provides a possible way to test the model versus observations: we know that the spectral index $-4$ for the PSD of the angular velocity can never be realised for any choice of the model's parameters in \eqref{P002_b}. 
{  On the other hand, observational evidence for a spectral index of $-4$ may be taken as an indication that gravitational radiation (which is not accounted for in the present model, but may be included as done in \citealt{meyers2021a}, see App.~\ref{app1}) is contributing to spinning down the pulsar.}

In general, a certain ordering of the corner frequencies gives rise to a sequence of spectral indexes in the PSD, 
\begin{equation}
\label{m2sequence}
    \begin{split}
        &\text{Possible case:} \quad  &\mu\, <\, \xi 
       \qquad  & ( \,-2,   \,\,\, 0 ,  -2 \,)
        \\ 
       &\text{Inconsistent case:}  \quad &   \xi \,<\, \mu
       \qquad & ( \,-2,  -4 , -2)
    \end{split}
\end{equation}
The underlying physical model tells us which spectral index sequence is consistent or inconsistent with it. The spectral sequences for the phase residuals are obtained from the ones above by adding $-2$ to each entry, e.g. $(-4,-2,-4)$ for the first case in \eqref{m2sequence}.
We can now use the noise parametrization in \eqref{infinitoso} and \eqref{psico} to write the corner frequencies as 
\begin{equation}
\label{corner_m2}
 \mu^2 \, = \,  \frac{ 1 }
 { 1+(x_1^2/x_p) (\alpha_\mathcal{T}/\alpha_\infty)^2 (\tau/T) } \cdot \dfrac{x_p^2}{\tau^2}
\, < \,  
\xi^2  \, = \, \dfrac{1}{\tau^2}  \, ,
\end{equation}
which tells us  that the region $\omega \in [x_p\,  \xi \, , \,  \xi]$ is always flat, whatever the choice of the parameters (this is, however, only the minimum possible extension of the flat region). 
In any case, to properly test the presence of this flattened frequency interval, the observed PSD should contain relevant information on the low-frequency domain, meaning that long observations (at least as long as a few times $\tau$) are needed. Moreover, according to our model pulsars with a bigger $\tau/T$ ratio are expected to display a more extended flat region. 
Below, we consider in more detail two interesting extreme cases. 
\\
\\
\emph{Pure external noise - } For $\alpha_{\mathcal{T}}=0$ we have $\xi=1/\tau$ and $\mu = x_p/\tau$: the flat region $\mu < \omega < \xi$ can not be less extended than this.  
The PSD in \eqref{P002} reduces to
\begin{equation}
\label{m2external}
    P_{p}(\omega)  
    = \frac{\alpha_\infty^2 \, \Omega \, \dot{\Omega}}{x_p^2 \, \omega^2} \cdot \frac{\omega^2+x_p^2\tau^{-2}}{\omega^2+\tau^{-2}}  \, \approx
    \frac{\alpha_\infty^2 \Omega \dot{\Omega}}{  \omega^2} \, ,
\end{equation}
where the approximation is valid only in the limit in which the loosely coupled superfluid component is restricted to the inner crust (for $x_p \approx 1$, we have $\mu \approx \xi$ and the flattened region disappears). 
As expected, in this limit we recover exactly the $m=0$ result in \eqref{PSD_L}.
\\
\\
\emph{Pure internal noise - } For $\alpha_\infty=0$ we have $\xi=1/\tau$ and $\mu = 0$, meaning that the flat region extends indefinitely at low frequencies.
The PSD in \eqref{P002} reduces to
\begin{equation}
\label{int2}
P_{p}(\omega)
  = \frac{ \alpha_\mathcal{T}^2 \,  x_1^2 \, \dot{\Omega}^2  }{2 \, x_p^2 \mathcal{B} \, \Omega} \cdot \frac{1}{\omega^2+\tau^{-2}} \, .
\end{equation}
This is the usual Lorentzian form of the PSD of a mean-reverting Ornstein-Uhlenbeck process, see Appendix \ref{appOU}.

\subsection{Spectral turnover for $\ensuremath{m=2}$}

For the three-component model $m=2$ the analysis of the PSD requires more care. 
Given the PSD in \eqref{P003_b}, we have 5 possibilities, which depend on the relative ordering of the four corner frequencies: following the scheme already used in \eqref{m2sequence}, we have that (only the possible cases that are consistent with \eqref{P003_b} are listed)
\begin{equation}
\label{2cases}
    \begin{split}
        \text{case 1:} \quad  &\mu_- \,<\, \mu_+ \,<\, \xi_- \,<\, \xi_+
         & ( \,-2,   \,\,\, 0 ,  +2, \,\,\,0, -2 \,)
        \\ 
       \text{case 2:}  \quad & \mu_- \,<\, \xi_- \,<\, \mu_+  \,<\, \xi_+
         & ( \,-2,  \,\,\,0 , -2, \,\,\,0, -2 \,)
        \end{split}
\end{equation}
For example, if the physical parameters of the $m=2$ model are such that case 1 is realised, then the expected sequence of spectral indexes is $(-2,0,+2, 0,-2)$, namely
\begin{equation}
\label{P003_case3}
  P_{p}(\omega) \propto 
 \begin{cases}
  \,\,  \mu_-^2 \mu_+^2 /(\xi_-^2 \xi_+^2 \, \omega^2)   
  \qquad & \omega \ll \mu_- 
  \\
  \,\,    \mu_+^2 / (\xi_-^2 \xi_+^2 )
  \qquad & \mu_- \ll \omega \ll \mu_+
  \\
  \,\,     \omega^2/ (\xi_-^2 \xi_+^2 \, ) 
  \qquad & \mu_+ \ll \omega \ll \xi_-
  \\
  \,\, 1/\xi_+^2
  \qquad & \xi_- \ll \omega \ll \xi_+
  \\
  \,\, 1/\omega^2 \qquad & \xi_+ \ll \omega 
  \end{cases}  
\end{equation}
Again, the spectral sequences for the phase residuals are obtained from the ones above by adding $-2$ to each entry.
Below, we consider in more detail two interesting extreme cases. With no loss of generality, we follow \citet{Pizzochero2020A&A} and set $0<b_1<b_2$: we will refer to the component 1 as ``loose'' superfluid component and to component 2 as  ``tight'' superfluid component (the loose component is the one that is less coupled to the normal one).
\\
\\
\emph{Pure external noise - } For $\alpha_{\mathcal{T}_1}=\alpha_{\mathcal{T}_2}=0$ we have that the two corner frequencies at the denominator $\xi_\pm$ are unchanged (they do not depend on the noise parameters), while 
\begin{equation}
    \mu_- = b_1 \qquad \mu_+ = b_2  \, .
\end{equation}
Moreover, it is possible to show that only case 2 in \eqref{2cases} can be realised, since
\begin{equation}
    b_1  < \xi_- < b_2 < \xi_+  \, .
\end{equation}
The PSD in \eqref{P003} boils down to 
\begin{equation}
    P_{p}(\omega) = \frac{ \alpha_\infty^2 \Omega \dot{\Omega} }{x_p^2 \, \omega^2} \cdot \frac{(\omega^2+b_1^2)(\omega^2+b_2^2)}{(\omega^2+\xi_-^2)(\omega^2+\xi_+^2) }\, .
\end{equation}
\\
\\
\emph{Pure internal noise in the ``loose'' component - } For $b_1<b_2$, we set $\alpha_\infty=\alpha_{\mathcal{T}_2}=0$, so that only the less coupled part of the superfluid is responsible for the fluctuations of the internal torque. This process is a Ornstein-Uhlenbeck process (see App. \ref{appOU}) and is asymptotically stationary.
Again,  only case 2 in \eqref{2cases} can be realised, since 
\begin{equation}
    \mu_- =0 \qquad \mu_+ = b_2  \, .
\end{equation}
and 
\begin{equation}
    0  < \xi_- < b_2 < \xi_+  \, .
\end{equation}
The PSD in \eqref{P003} reads, cf. with \eqref{int2},
\begin{equation}
    P_{p}(\omega) = \frac{ \alpha_{\mathcal{T}_1}^2 \Omega \dot{\Omega} }{x_p^2 \, \omega^2}
    \cdot \frac{x_1^2}{b_1 T}
    \cdot \frac{ \omega^2 (\omega^2+b_2^2)}{(\omega^2+\xi_-^2)(\omega^2+\xi_+^2) }\, .
\end{equation}
Recalling that $b_i = 2 \Omega \mathcal{B}_i$, the PSD can also be written as
\begin{equation}
    P_{p}(\omega) = \frac{ \alpha_{\mathcal{T}_1}^2  \dot{\Omega}^2 }{x_p^2  }
    \cdot \frac{x_1^2}{2 \Omega \mathcal{B}_1  }
    \cdot \frac{   (\omega^2+b_2^2)}{(\omega^2+\xi_-^2)(\omega^2+\xi_+^2) }\, .
\end{equation}
\\
\\
\emph{Pure internal noise in the ``tight'' component - } For $b_1<b_2$, we set $\alpha_\infty=\alpha_{\mathcal{T}_1}=0$, so that only the more tightly coupled part of the superfluid is responsible for the fluctuations of the internal torque. Again, this process is a Ornstein-Uhlenbeck process (see App. \ref{appOU}) and is asymptotically stationary.
Now, only case 1 in \eqref{2cases} can be realised, since 
\begin{equation}
    \mu_- =0 \qquad \mu_+ = b_1  \, .
\end{equation}
and 
\begin{equation}
    0  < b_1 < \xi_- < \xi_+  \, .
\end{equation}
The PSD in \eqref{P003} reads, cf. with \eqref{int2},
\begin{equation}
    P_{p}(\omega) = \frac{ \alpha_{\mathcal{T}_2}^2 \Omega \dot{\Omega} }{x_p^2 \, \omega^2}
    \cdot \frac{x_2^2}{b_2 T}
    \cdot \frac{ \omega^2 (\omega^2+b_1^2)}{(\omega^2+\xi_-^2)(\omega^2+\xi_+^2) }\, .
\end{equation}
and we can have a region of the spectrum $b_1<\omega<\xi_-$ that is blue~(spectral index $+2$ for the angular velocity).

\section{Timing noise strength}
\label{sec:tnstr}

Several metrics have been proposed to quantify the strength  of observed timing noise in a certain pulsar. To extract the timing noise strength from our simulated signal $\Omega_p$, we have to mimic the observational procedure.
First, we introduce the total phase difference of the observable signal between two arbitrary times $0$ and $t$ in the usual way, 
\begin{equation}
\label{ang_phase}
     {\phi}_t = \int_0^t {\Omega_p}(z) \,  dz  \, .
\end{equation}
In timing observations, the phase residuals (i.e., the observational counterpart of our theoretical process {  $\delta \phi_t$}) are measured with respect to a given timing model with phase $\mathbf{\phi}^{\text{mod}}_t$, typically modelled as a power series in $t$ \citep{Lorimer_book_2004}, namely
\begin{equation}
\label{ang_phase_res}
     \delta {\phi}_t =  {\phi}_t  -  {\phi}^{\text{mod}}_t  \, 
     \qquad 
     \dot{\phi}^{\text{mod}}_t  = \Omega^{\text{mod}}  + \dot{\Omega}^{\text{mod}} \, t + ... 
\end{equation}
In our case, to extract the theoretical residuals we can set,  
\begin{equation}
    \label{phase_model}
     \phi^{\text{mod}}_t = \Omega_p(0) \, t - |\dot{\Omega}_\infty| \, t^2/2 \, , 
\end{equation}
or, in terms of angular velocity,
\begin{equation}
    \label{baule}
    \delta\Omega_p = \Omega_p(t) - \dot{\phi}^{\text{mod}}_t 
    \qquad \qquad 
    \langle \Omega_p(t) \,\rangle\, = \, \dot{\phi}^{\text{mod}}_t \, .
\end{equation}
Therefore, the biggest difference between observations and our theoretical setting is that in real observations the timing model ${\phi}^{\text{mod}}_t $ is derived from the prior knowledge of the rotational ($\Omega$ , $\dot\Omega$) and astrometric parameters. We may follow the same procedure and consider a more general $ \phi^{\text{mod}}_t $ that does not necessarily coincide with the deterministic solution ($\Omega_p(0)$ and $|\dot{\Omega}_\infty|$ are just the input of a mathematical model and can not be extracted directly from observations), namely $\phi^{\text{mod}}_t = \phi_0 + \Omega \, t - |\dot{\Omega}| \, t^2/2$, with 
$\Omega \approx \Omega_p(0)$ and $\dot{\Omega} \approx |\dot{\Omega}_\infty|$ inferred directly from the simulated process over a finite time window.
However, this complicates considerably the following calculations, so we stick to the prescriptions in \eqref{phase_model} and \eqref{baule}. Therefore, our analysis refers to the ideal case in which the long-term spin-down rate $\dot{\Omega}$ extracted from a fit to the data well approximates the intrinsic (not directly observable) parameter $|\dot{\Omega}_\infty|$ that sets the average spin-down rate of the pulsar over many decades.

\subsection{Reconstructing the times of arrival}
\label{sec_times}

Numerical simulations of pulsar rotation are based on evolving some (simplified) dynamical equations for the angular velocity, which is then obtained directly by numerical integration of the assumed theoretical model. However, in pulsar timing analysis, the quantity that is measured directly are the TOAs of a train of pulses, while the phase $\phi(t)$ and the angular velocity are derived quantities.
Therefore, if we want to study the residuals of the TOAs with respect to a certain timing model $\phi^{\text{mod}}(t)$, we first have to reconstruct them. 

Let us indicate the measured pulsar's TOAs by $\{t_j\}$, while the TOAs of the reference polynomial model are $\{t^\text{mod}_j\}$, meaning that each $t_j$ and $t^\text{mod}_j$ are the times such that the pulsar's phase $\phi\left( t_j \right)$ coincides with the one of the model\footnote{
    From the observational point of view,  a fiducial point in the observed pulse profile is chosen a priori to keep track of the pulsar's phase. The pulsar must have spun around by an integer number of rotations $ \Delta N$ between two measured TOAs $t_j$ and $t_{j+1}$, provided there is no change in the assumed timing model. 
    Therefore, \eqref{fundamental_thing} is valid because according to the observational procedure used to extract the TOAs, 
    $\phi(t_{j+1}) - \phi(t_{j}) = 2 \pi \Delta N = \phi^{\text{mod}}(t^{\text{mod}}_{j+1}) - \phi(t^{\text{mod}}_{j})$ and we are free to set the phase relative to the first TOA equal to zero.
}
\begin{equation}
\label{fundamental_thing}
    \phi\left( t_j \right)  = \phi^\text{mod}\left( t^\text{mod}_j \right)   \, .
\end{equation}
In terms of the timing residuals 
\begin{equation}
\label{daj}
    \delta a_j = t_j - t^\text{mod}_j   \, ,
\end{equation}
that measure how much the observed TOAs advance or lag behind the ones expected for the timing model, the above equation reads 
\begin{equation}
    \phi(t^\text{mod}_j + \delta a_j)=\phi^\text{mod}(t^\text{mod}_j) \, .
\end{equation}
We can now generalise this concept to a continuous observation in time by demanding that $\delta a_t$ is the solution of the implicit equation
\begin{equation}
    \phi(t + \delta a_t ) = \phi^\text{mod}(t) \, .  
\end{equation}
Implicit differentiation of the above equation gives
\begin{equation}
\label{dini}
    \dfrac{d}{dt}\, \delta a_t = - \dfrac{ \Omega_p(t) - \dot{{\phi}}^\text{mod}_t }{ \Omega_p(t)}
    -\dfrac{\dot{{\phi}}^\text{mod}_t \dot{\Omega}_p(t)}{ { \Omega_p(t)}^2} \, \delta a_t + O\left( \dfrac{\delta a_t^2}{T^2}\right ) \, ,
\end{equation}
where $T$ is of the order of the pulsar's age. 
{  On timescales much smaller than the pulsar age, we can approximate $\Omega_p$ with a suitable constant value $\Omega$, so that the expression 
\begin{equation}
\label{semplice}
    \delta a_t \, \approx \,- \,\dfrac{ \delta\phi(t)}{\Omega} 
\end{equation}
is (for our purposes) a very good approximation to the exact solution of \eqref{dini} for time intervals much shorter than $T$: } 
while \eqref{dini} leads to a more accurate reconstruction of $\delta a_t$, for the sake of studying the timing noise strength $\sigma^2$ {  (to be introduced in the next subsection)}, the simple prescription \eqref{semplice} turns out to be good enough.

\subsection{Timing noise strength}

The  variance $\sigma^2$ of the timing residuals from a least squares polynomial fit over an observation interval of length $T_o$ provides a measure of the timing noise strength in real timing data \citep{Cordes1980_III,hobbs_2010MNRAS,Shannon_noiseMSP_2010}:
\begin{equation}
\label{sigma_obs}
    \sigma^2 = \frac{1}{N_o} \sum_{j=1}^{N_o} \, \left| \delta a_j  \right|^2
     \, , 
\end{equation}
where $\delta a_j$ are the measured timing residuals in \eqref{daj} and $N_o$ is the number of measurements. For a continuous-time model we can express the above discrete expression in terms of the variance of the continuous process $\delta a_t$, 
\begin{equation}
\label{sigma_continuo}
    \sigma^2 = \frac{1}{T_o} \int_{t_0}^{t_0+T_o} \!\!dz\,\left| \delta a_z \right|^2 \, 
    \approx 
    \frac{1}{T_o \, \Omega^2} \int_{t_0}^{t_0+T_o}\!\! dz\,\left|\delta \phi_z \right|^2
     \, , 
\end{equation}
where the timing residuals $\delta a_t$ have been translated into phase residuals (that are more easily obtained in numerical simulations) via \eqref{semplice}. 
In this way, $\sigma$ is the  {root mean square} of the observed phase residuals ${\phi}_j$ and its physical dimension is that of time \citep{Cordes1980_III}.
To estimate $\sigma$ from our model, we follow the same procedure that we have used to obtain the PSD for the angular velocity. 
First, we introduce the PSD for the phase residuals and for the TOA residuals as 
\begin{equation}
\label{pphi1}
    P_\phi (\omega) = \omega^{-2} \, P_p (\omega) 
    \qquad \quad 
    P_a (\omega) \approx (\omega \Omega)^{-2} \, P_p (\omega)\,  ,
\end{equation}
where the expression for the PSD of the TOAs residuals $ P_a (\omega) $ is an approximation based on \eqref{semplice}. 
We may estimate $\sigma^2$ by considering the total power contained in a band of frequencies $\omega \in [\nu_o , \nu_s]$, namely 
\begin{equation}
\label{quellochevoglio}
    \sigma^2 \approx 2 \int_{\nu_o}^{\nu_s} \dfrac{d\omega}{2 \pi} \, P_a(\omega)
     \approx\frac{2}{\Omega^2} \int_{\nu_o}^{\nu_s} \dfrac{d\omega}{2 \pi} \, P_\phi(\omega) 
    \, .
\end{equation}
The above integral is band-limited because (in observations as well as in simulations) we can not access frequencies higher than the sampling one  $\nu_s \sim N_o/T_o$. Moreover, the lower frequency $\nu_0 \sim 1/T_0$ acts as a cut-off parameter that regulates the  divergence of $P_\phi(\omega)$ for $\omega \rightarrow 0$. The prefactor of $2$ is needed since we are integrating over positive frequencies only, and the overall scaling with $\Omega^2$ comes from equation \eqref{semplice}.

The expression for $\sigma^2$ in  \eqref{quellochevoglio} is a guess and it is not equivalent to \eqref{sigma_continuo}; however, the scaling of $\sigma^2$ with the physical and rotational parameters turn out to be the same, whether we use \eqref{quellochevoglio} or \eqref{sigma_continuo}: the two estimates of $\sigma^2$, can  differ by a dimensionless constant of order 1 (not surprisingly, given the arbitrariness of the cut-off $\nu_0$). 
We analytically check this claim  in App. \ref{appOU}.

\subsection{Timing noise strength for $m=1$ and $m=2$}

In order to see if our model can reproduce the observed scaling of $\sigma^2$ reported in the literature, e.g. \citep{hobbs_2010MNRAS,Shannon_noiseMSP_2010,Parthasarathy2020MNRAS_I} we explicitly calculate the band-limited integral in \eqref{quellochevoglio} for the $m=1$ model.

The timing noise strength is often studied across a sample of pulsars. 
For example, \citet{Cordes1980_III} found that, out of a sample of 11 pulsars, the timing noise strength is correlated with the period derivative and weakly with the period. 
More precisely, analysis of the timing noise strength is based on the study of correlations in the data couples $(\, \Omega^{a}_i \dot\Omega^{ b}_i \, , \, \sigma_i  \, )$, where $\Omega_i$, $\dot\Omega_i$ and $\sigma_i$ are observed properties of a given $i$-th pulsar. 
Following \citet{hobbs_2010MNRAS}, multiple combinations of $(a,b)$ can be tested to maximise some measure of correlation (like the Pearson coefficient), which leads them to find that  $\sigma \propto -1.37 \log_{10}[\Omega^{0.29}  \dot\Omega^{-0.55}]$, i.e.  $a\approx-0.39$, $b\approx0.75$. 
They also conclude that timing noise is inversely correlated with characteristic age: in fact, the phenomenological scaling of $\sigma$ can also be expressed as
\begin{equation}
\label{scaling_T}
    \sigma \propto \Omega^a \dot\Omega^b = \Omega^{a+b} T^{-b} =  T^{a} \dot\Omega^{a+b} \, ,
\end{equation}
implying that we can equivalently interpret the result of Hobbs and collaborators as $\sigma \propto P^{-0.36} T^{-0.75}$, where $P \propto \Omega^{-1}$ is the pulsar's period.

More recently, \citet{Shannon_noiseMSP_2010} reported an observational scaling relation of $ \sigma \propto \Omega^{-0.9\pm 0.2} \dot\Omega^{1 \pm 0.05}$ for a sample of canonical pulsars (CPs) and $a \approx-1.5$,  $b \approx 1.2$ for the whole pulsar population.
This relationship has also been investigated by \citet{Parthasarathy2020MNRAS_I}; they found that $ \sigma \propto \Omega^{1} \dot\Omega^{-0.9\pm0.2 }$ for the sample of young pulsars, whereas after adding a few millisecond pulsars (MSPs) they found that $ \sigma \propto \Omega^{1} \dot\Omega^{-0.6\pm0.1 }$ on the basis of an analysis based on the Spearman's rank correlation coefficient. 
However, this kind of statistical analysis can only probe the ratio $b/a$ of $\Omega^a \dot\Omega^b $ (see \citet{Jankowski}). In fact,  any rescaling $\Omega^{ca} \dot\Omega^{cb} $ with $c>0$ preserves the ranking of the data: the only thing that the Spearman's coefficient can probe is the degree of monotonic behaviour (i.e., the ranking) of the observed data couples $( \Omega^{1}_i \dot\Omega^{ b/a}_i , \sigma_i  )$, where $i$ indicates a particular pulsar.
Hence, the relationship found by \citet{Parthasarathy2020MNRAS_I}, $b/a\approx -0.6 $, is only qualitatively consistent with the previous result of  \citet{Shannon_noiseMSP_2010} for a sample of both canonical pulsars (CP) and millisecond pulsars (MSP), $ b/a \approx -0.8$, see Table \ref{tab_sigma}. 
{  Finally, \citet{Lower_Utmost_II} have developed a Bayesian framework to estimate the scaling parameters $a$ and $b$. Their result (see Tab. 6 in their original work),  is also listed in Tab. \ref{tab_sigma} and is particularly close to the earlier result of \citet{Shannon_noiseMSP_2010} for non-recycled pulsars (CP).  } 

We now provide a theoretical estimate of the timing noise strength $\sigma$ and its scaling parameters $a$ and $b$ by assuming that timing noise follows our stochastic model. 
In principle, it is possible to obtain the exact analytical expression for \eqref{quellochevoglio} by plugging the PSD \eqref{P002} into \eqref{pphi1}, but the final expression is quite convoluted. 
Therefore, we only consider the two limiting cases of pure internal and pure external noise. 
To find the dominant dependence of $\sigma$ on the rotational parameters,  we assume that there is a hierarchy of timescales, namely 
\begin{equation}
\label{gerarchia}
     T_o/N_o \ll \tau \ll T_o \ll  T \quad  \quad 
     \left(\, \dot{\Omega} \ll \Omega \, \nu_o 
     \ll \mathcal{B}\,\Omega^2 \ll \Omega \, \nu_s \, \right)
\end{equation}
The explicit calculations for $m=1$ are done by considering the results in  \eqref{P002} and the formula for $\sigma^2$ in \eqref{quellochevoglio}.
\\
\\
\emph{Pure external noise  ($m=1$) -} Plugging the exact PSD in \eqref{m2external} into our definition \eqref{quellochevoglio}, we obtain the leading terms
\begin{equation}
\label{phi2external}
\sigma^2 \approx
\dfrac{\alpha_\infty^2  \dot\Omega}{3 \pi \nu_o^3  \Omega}
   \left( 1+ \dfrac{3(1-x_p^2)\nu_o^2}{4 \mathcal{B}^2 \Omega^2}\right)   
\end{equation}
To obtain this result we have expanded the exact result to implement the hierarchy of timescales in \eqref{gerarchia}. The leading term tell us that $\sigma^2 \propto \Omega^{-1} \dot\Omega$. 
\\
\\
\emph{Pure internal noise ($m=1$) - } We now plug the exact PSD in \eqref{int2} into  \eqref{quellochevoglio} and expand according to the hierarchy of timescales in \eqref{gerarchia}:
\begin{equation}
\label{phi2internal}
\sigma^2 \approx
\dfrac{\alpha_\mathcal{T}^2 \, x_1^2 \, \dot\Omega^2}
   {8 \pi \, \nu_o  \,\mathcal{B}^3 \, \Omega^5 }
   \left( 1- \dfrac{ \pi\,x_p \, \nu_o  }{  4 \,\mathcal{B} \, \Omega   }\right)
\end{equation}
Hence, the expected noise strength is expected to scale with the rotational parameters as $\sigma^2 \propto \Omega^{-5} \dot\Omega^2$.
\\
\\ 
For $m=2$, the calculation proceeds along the same steps as previously, albeit it is more laborious. Again, we take component 1 to be the loose one, so that the analogous of \eqref{gerarchia} is 
\begin{equation}
\label{gerarchia2}
    \dot{\Omega} \ll \Omega \, \nu_o 
     \ll \mathcal{B}_1\,\Omega^2 \ll \mathcal{B}_2\,\Omega^2 \ll \Omega \, \nu_s \, .
\end{equation}
We perform the integral in \eqref{quellochevoglio} by using the relations in \eqref{P003} and then expand the result thanks to the hierarchy of timescales in the above equation; the final result reads:
\begin{equation}
    \begin{split}
    & \sigma^2 \approx
        \dfrac{\alpha_\infty^2  \dot\Omega}{3 \pi \nu_o^3  \Omega}
   \left( 1+ \dfrac{3(2-x_1) x_1  \, \nu_o^2}{4 \,\mathcal{B}^2_1 \, \Omega^2}\right)
        \quad  & (\alpha_{\infty}\neq 0)
        \\
    &    \sigma^2 \approx
\dfrac{\alpha_{\mathcal{T}1}^2 \, x_1^2 \, \dot\Omega^2}
   {8 \pi \, \nu_o  \,\mathcal{B}_1^3 \, \Omega^5 }
   \left( 1- \dfrac{ \pi\,(1-x_1) \, \nu_o  }{  4 \,\mathcal{B}_1 \, \Omega   }\right)
     \quad  & (\alpha_{\mathcal{T}1} \neq0) 
        \\
    &    \sigma^2 \approx
\dfrac{\alpha_{\mathcal{T}2}^2 \, x_2^2 \, \dot\Omega^2}
   {8 \pi \, \nu_o  \,\mathcal{B}_2^3 \, \Omega^5 }
   \left( 1 + \dfrac{ \pi\,(2-x_1) x_1\, \nu_o  }{  4 (1-x_1)\,\mathcal{B}_1 \, \Omega   }\right)
        \quad  & (\alpha_{\mathcal{T}2} \neq0) 
    \end{split}
\end{equation}
for the pure external noise ($\alpha_{\mathcal{T}1},\alpha_{\mathcal{T}2}=0$), noise in the loose component ($\alpha_{\infty},\alpha_{\mathcal{T}2}=0$) and noise in the tight component ($\alpha_{\mathcal{T}1},\alpha_{\infty}=0$), respectively. The leading terms are identical to the corresponding ones for the $m=1$ cases. The sub-leading terms have different dimensionless parameters, but they still obey the same scaling of the $m=1$ model. 
In the end, recalling that $\nu_0 \sim 1/T_0$, we can conclude that, for both $m=1,2$,
\begin{equation}
\label{faticafinale}
    \begin{split}
    & \sigma^2 \approx
    a_{e0} \, \dot{\Omega}  \Omega^{-1} T_0^3 + a_{e1} \, \dot{\Omega}  \Omega^{-3} T_o
        \quad  & (\text{pure external noise})
        \\
    & \sigma^2 \approx
    a_{i0} \, \dot{\Omega}^2  \Omega^{-5} T_o  + a_{i1} \, \dot{\Omega}^2 \Omega^{-6}      
    \quad  & (\text{pure internal noise})
    \end{split}
\end{equation}
where $|a_{e1}| \ll a_{e0}$ and $|a_{i1}| \ll a_{i0}$ are dimensionless constants that depend on the physical parameters of the pulsar. 
The results are summarised in Table \ref{tab_sigma}. {  We see that the stochastic model for pure external noise captures the value of the ratio $b/a$, but their individual values are rather different from the recent Bayesian estimates by \citet{Lower_Utmost_II}. Therefore, we are left with three possibilities. The first one is that a model with pure external torque is not good enough to reproduce the scaling and that some amount of internal fluctuations are necessary (adding internal fluctuations raises the value of $b$ and decreases $a$, bringing their values closer to observed ones). The second possibility is that a model with injected white noise should be abandoned in favour of coloured noise (noise due to internal turbulence is red, see e.g. \citealt{MelatosLink2014}).
Finally, it is also possible that our parameters $\alpha_{\infty,\mathcal{T}}$ depend on the pulsar's age, so that they carry extra correlation with $\Omega$ and $\dot{\Omega}$ that is unaccounted here. }

\begin{table}
\begin{center}
\begin{tabular}{ c|c|c|c } 
  $\sigma \propto \Omega^a \dot\Omega^b$ & $a$ & $b$ & $b/a$ \\
 \hline
``external'' noise, eq. \eqref{faticafinale} & -0.5  &  0.5 &  -1   \\ 
``internal'' noise, eq. \eqref{faticafinale} & -2.5  &  1   &  -0.4 \\ 
 \citet{Lower2021mnras} & -0.84  & 0.97   & -1.15 \\
 \citet{Parthasarathy2020MNRAS_I} & 1 (assumed) & -0.6 & -0.6 \\ 
 \citet{Shannon_noiseMSP_2010} (MSP+CP) &  -1.5 &  1.2 &  -0.8 \\
 \citet{Shannon_noiseMSP_2010} (CP) & -0.9 &1 & -1.1 \\
 \citet{hobbs_2010MNRAS} &   -0.39  &   0.75  &   -1.9 \\
 \hline
\end{tabular}
\end{center}
\caption{
    Scaling of the timing noise strength $\sigma$ for our model (considering the leading terms in \eqref{faticafinale}, that are valid for both the $m=1$ and $m=2$ models) and extracted from pulsar timing observations. The result of \citet{Parthasarathy2020MNRAS_I} is closer to the case of pure internal noise, while the analysis of \citet{Shannon_noiseMSP_2010} returns a ratio $b/a$ that is consistent with the one of pure external noise for canonical pulsars (CP).  The values of \citet{hobbs_2010MNRAS} are obtained from their equation (6) after multiplying both $a$ and $b$ by their prefactor $-1.37$. For notation convenience, $\dot{\Omega}>0$ is the absolute value of the spin down rate throughout the paper.
    }\label{tab_sigma} 
\end{table}

\begin{figure*}
    \centering
    \includegraphics[scale=0.049]{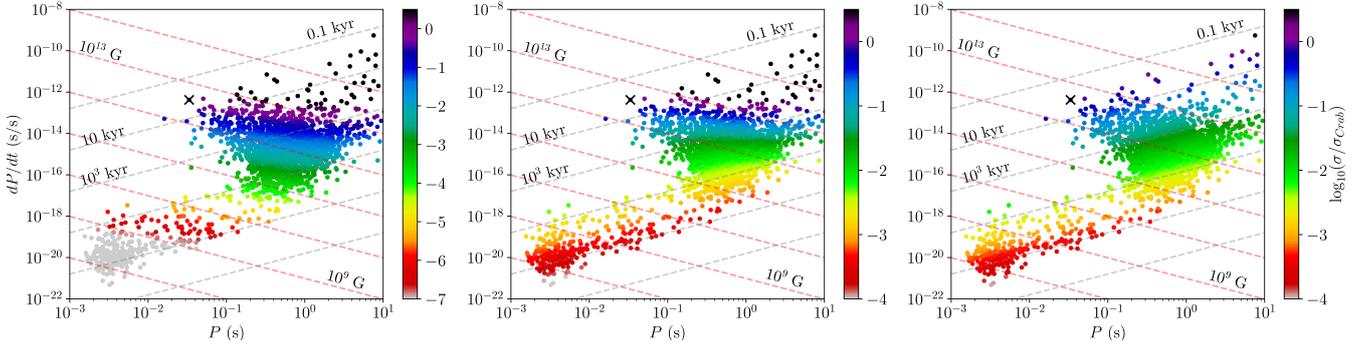}
    \caption{
    The $P-\dot{P}$ diagram for all pulsars listed in the ATNF catalogue (accessed on 2022 April 1). The colour bar represents the theoretically predicted timing noise strength $\sigma$ for $m=1$ on a logarithmic scale, normalised over the Crab's value of $\sigma$. 
    In all three panels the common parameters are: $\mathcal{B}=2\times 10^-9$, $x_1=0.1$, $\nu_{o,s} = 2 \pi/T_{o,s}$, where $T_o=20\,$yr and $T_s=1$ day. 
    The cross marks the Crab pulsar (PSR J0534+2200). Left: pure internal noise for $\alpha_\mathcal{T} =10^{-2}$, the timing noise strength approximately scales as $\sigma \propto \dot{P} P^{0.5}$. Centre: $\alpha_\mathcal{T} =10^{-2}$, $\alpha_\infty =10^{-7}$. Right: pure external noise for $\alpha_\infty =10^{-7}$, the timing noise strength approximately scales as $\sigma \propto \sqrt{\dot{P}/P} \propto T^{-0.5}$.
    The lines of constant magnetic field for $10^8-10^{14}\,$G (red, dashed) and the lines of constant characteristic age $T/2=0.1-10^9\,$kyr (grey, dashed) are also indicated to guide the eye. 
        }\label{figppdot}
\end{figure*}

Keeping in mind the three possibilities above, we stick to our minimal setting and tentatively explore how our results for the timing noise strength $\sigma$ vary across the known pulsar population. 
This is done in Fig. \ref{figppdot}, where each pulsar in the $P-\dot{P}$ diagram is plotted according to a colour scheme representing the theoretical expectation for its timing noise strength: for each object we use the exact formula \eqref{quellochevoglio} for $m=1$ and $\mathcal{B}=2\times 10^-9$, $x_1=0.1$, $\nu_{o,s} = 2 \pi/T_{o,s}$, where the observation time is $T_o=20\,$yr and the sampling time is $T_s=1$ day. In the case of pure external noise (right panel) the expected scaling is $\sigma \propto 1/\sqrt{T}$, in accordance with the leading term of the first equation in \eqref{faticafinale}.

 We see that fluctuations in the internal torque tend to define the scaling in the upper part of the diagram (the one containing younger pulsars). On the other hand, even a small amount of noise in the external torque dominates in the region of older pulsars, as can be seen in the central panel. 
 
 The use of the exact formula \eqref{quellochevoglio} results in corrections that do not modify the general approximate scaling defined by the leading terms in \eqref{faticafinale}, at least in the region $P$ and $\dot{P}$ of known pulsars. 
 In fact, equation \eqref{faticafinale} tells us that, in the case of pure external noise, the leading term is $\sigma \propto T^{-0.5}$, which happens to be the case in the right panel of Fig. \ref{figppdot}, where the lines of constant $\sigma$ coincide with the lines of constant age.
 On the other hand, for pure internal noise, the approximate scaling in \eqref{faticafinale} can be written as $\sigma \propto P^{1/2}\dot{P} $, so that the lines of constant $\sigma$ should have slope $-0.5$ (to be compared with the slope $-1$ of the lines of constant magnetic field). Again, this is what can be seen in the left panel of Fig. \ref{figppdot}: the full result \eqref{quellochevoglio} is in good agreement with the approximate scaling defined by the leading term in \eqref{quellochevoglio}.
 
 Clearly, these results for the pulsar population are biased by the fact that we have set constant values of $\mathcal{B}$, $x_1$, $\alpha_\infty$ and  $\alpha_\mathcal{T}$ across the whole population. Actually, these physical parameters are likely to have an intrinsic scaling with the pulsar age (e.g., mature pulsars may have a smaller value of $\mathcal{B}$, or a larger $x_1$, than young ones), implying  that the coefficients $a_{0,1}$ in \eqref{faticafinale} are also functions of $\Omega$ and $\dot \Omega$ via their combination $T$. This also affects the theoretical values of $a$ and $b$ in Tab. \ref{tab_sigma}.


\section{Numerical tests}
\label{sec:num}


The It\^{o}'s process in \eqref{ito} may be integrated forward in time thanks to the simple Euler–Maruyama scheme, that is often used to solve the Ornstein–Uhlenbeck process (see App. \ref{appOU}); for a finite time step of amplitude $\Delta t>0$, we have 
\begin{equation}
\label{em_scheme}
    \delta\mathbf{\Omega}_{t+\Delta t }
    = \delta\mathbf{\Omega}_t+ B \,\delta\mathbf{\Omega}_{t } \, \Delta t + M \,\mathbf{X}\, \sqrt{\Delta t} + O(\Delta t^{3/2})
\end{equation}
where $\mathbf{X}=(\mathbf{W}_{t+\Delta t }-\mathbf{W}_{t })/\sqrt{\Delta t}$ is a vector of i.i.d. normal random variables sampled at every time step.
To give an explicit example, for the $m=1$ system defined by \eqref{2_comp_mat} and \eqref{M2}, the above scheme reads  {  (the shorthand $\Omega(t)=\Omega^t$ is used):}
\begin{equation}
\label{scheme_m2}
    \begin{split}
   & \delta{\Omega}_p^{t+\Delta t}
    = \delta{\Omega}_p^{t} + x_1\frac{\delta{\Omega}_1^{t}-\delta{\Omega}_p^{t}}{\tau} \, \Delta t 
    +   \left[ X_p \frac{\sigma_\infty}{x_p} - X_1 \frac{\sigma_\mathcal{T}}{x_p} 
     \right]  \sqrt{\Delta t}
   \\
    & \delta{\Omega}_1^{t+\Delta t}
    = \delta{\Omega}_1^{t} - x_p\frac{\delta{\Omega}_1^{t}-\delta{\Omega}_p^{t}}{\tau} \, \Delta t 
    + X_1 \frac{\sigma_\mathcal{T}}{x_1} \sqrt{\Delta t} 
   \\
   &  X_p \sim \mathcal{N}(0,1)\, , \quad  X_1 \sim \mathcal{N}(0,1) 
   \end{split}
\end{equation}
where $\sigma_\infty$ and $\sigma_\mathcal{T}$ are given in \eqref{infinitoso} and \eqref{psico}, respectively. In the end, the evolution in \eqref{scheme_m2} will depend on six physical parameters:  $x_1$, $\alpha_\infty$, $\alpha_\mathcal{T}$, $\mathcal{B}$, $\Omega$, $\dot{\Omega}$. 
Moreover, the scheme in \eqref{scheme_m2} is consistent with the evolution of the angular momentum residuals\footnote{
    Recall that $\dot\Omega >0$ is just a benchmark value for the observed spin-down rate $|\dot{\Omega}_p|$, to be set from observations together with $\Omega>0$.
}
\begin{equation}
\label{scheme_L}
     \delta{L}_{t+\Delta t}
    = \delta{L}_{t}  + X_0 \, \sigma_\infty \, \sqrt{\Delta t} =
    \delta{L}_{t}  + X_0 \, \alpha_\infty   \sqrt{\Omega \, \dot{\Omega} \, \Delta t}
    \, .
\end{equation}
The $m=2$ case in \eqref{m3B} and \eqref{m3M} is completely analogous, with $\sigma_\infty$ that is still set according to \eqref{infinitoso}, while the two parameters $\sigma_{\mathcal{T}_1}$ and $\sigma_{\mathcal{T}_2}$ are given in \eqref{psico3}. In order to capture the features of the PSD and to ensure that the scheme is linearly stable, the ``sampling rate'' $\nu_s=1/\Delta t$ is always chosen to be considerably higher than the largest eigenvalue of $-B$. For example, we choose $\Delta t \ll b^{-1}$ for $m=1$ and $\Delta t \ll \min(b_1^{-1},b_2^{-1})$ for $m=2$. Note that $b_1$ and $b_2$ are not eigenvalues of $B$ for $m=2$, but this condition automatically ensures that the linear stability criterion is satisfied, see equation (A.19) of \citealt{montoli2020A&A}.

\subsection{Evolution of a Vela-like pulsar}

In order to build some intuition about the phenomenology associated with the stochastic process introduced in Section 1, we show the results of some simple numerical tests for a pulsar that has the rotational parameters of the Vela (PSR J0835-4510, see e.g. \citealt{ATNFcatalogue}); we set $\Omega_p(0) = 70\,$rad/s, $|\dot\Omega_\infty|= 10^{-10}\,$rad/s$^2$. We advance the equations in \eqref{scheme_m2} for $20\,$yr, roughly the span of the phase-coherent timing solution for the Vela that encompasses $\sim 10^3$ TOAs obtained with the Parkes radio telescope (roughly) between 1993 and 2014 \citep{shannon_21yr_vela}. 

\begin{figure}
    \centering
    \includegraphics[scale=0.45]{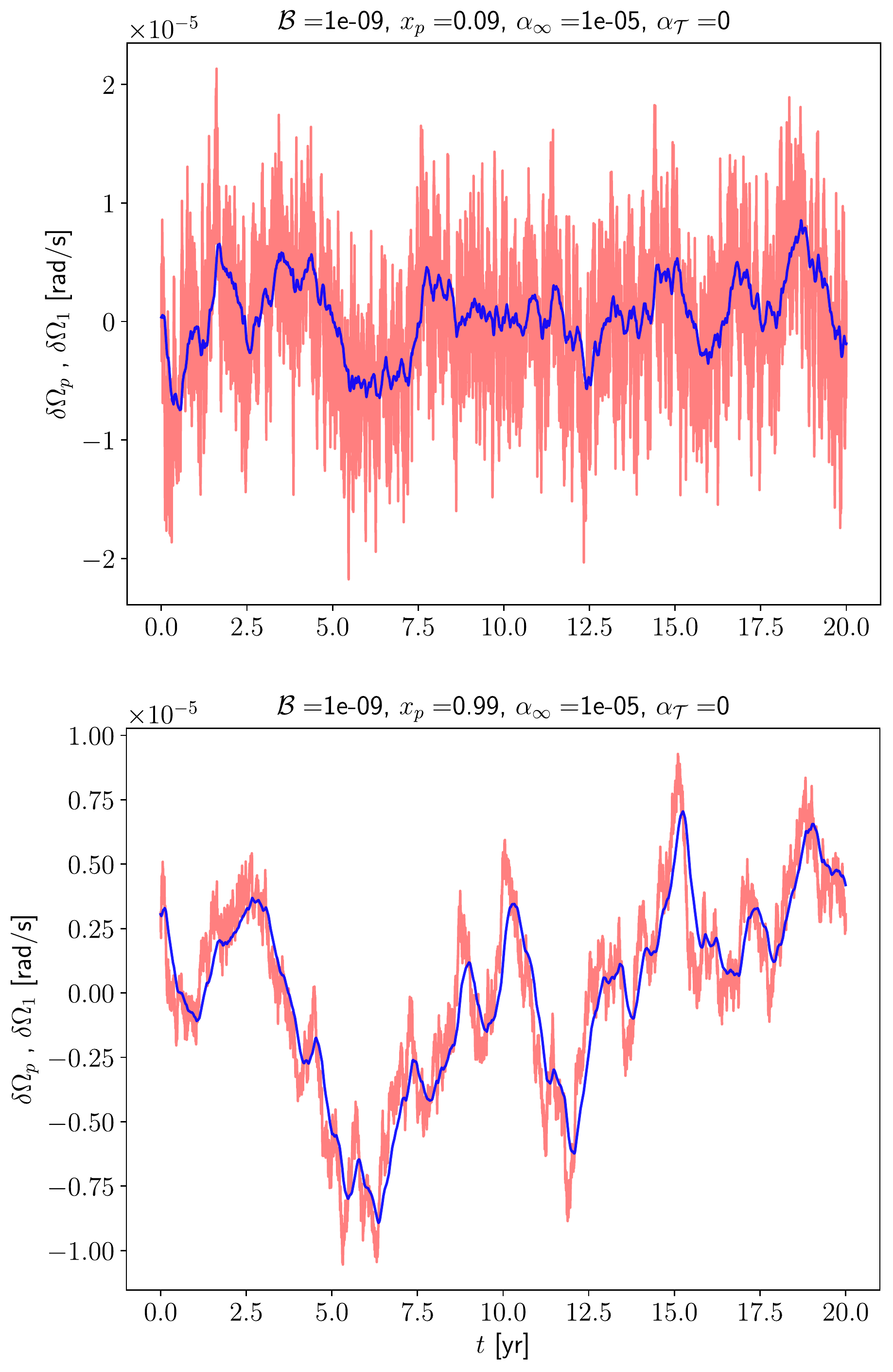}
    \caption{
    A typical realisation of the processes $\delta\Omega_p$ (pink) and $\delta\Omega_1$ (blue) for the two component model ($m=1$) with pure external noise ($\alpha_\infty=10^{-5}$, $\alpha_\mathcal{T}=0$), over a time span of 20 years. The rotational parameters are the ones of the Vela pulsar, $\Omega\approx 70\,$rad/s, $\dot\Omega\approx 10^{-10}\,$rad/s$^2$. In both cases $\mathcal{B}=10^{-9}$ and the only difference between the two panels is the value of the superfluid ratio: $x_p=0.1$ in the upper panel (the star is mostly superfluid and the resulting relaxation timescale is $\tau \approx 8\,$d), $x_p=0.99$ in the lower panel (the superfluid is confined in the inner crust, $\tau \approx 80\,$d). The fluctuations in the external torque drive the observable component, while the superfluid tends to follow it with some delay (this is even more evident in Fig. \ref{figVela2}). 
    }
    \label{figVela1}
\end{figure}

\begin{figure}
    \centering
    \includegraphics[scale=0.45]{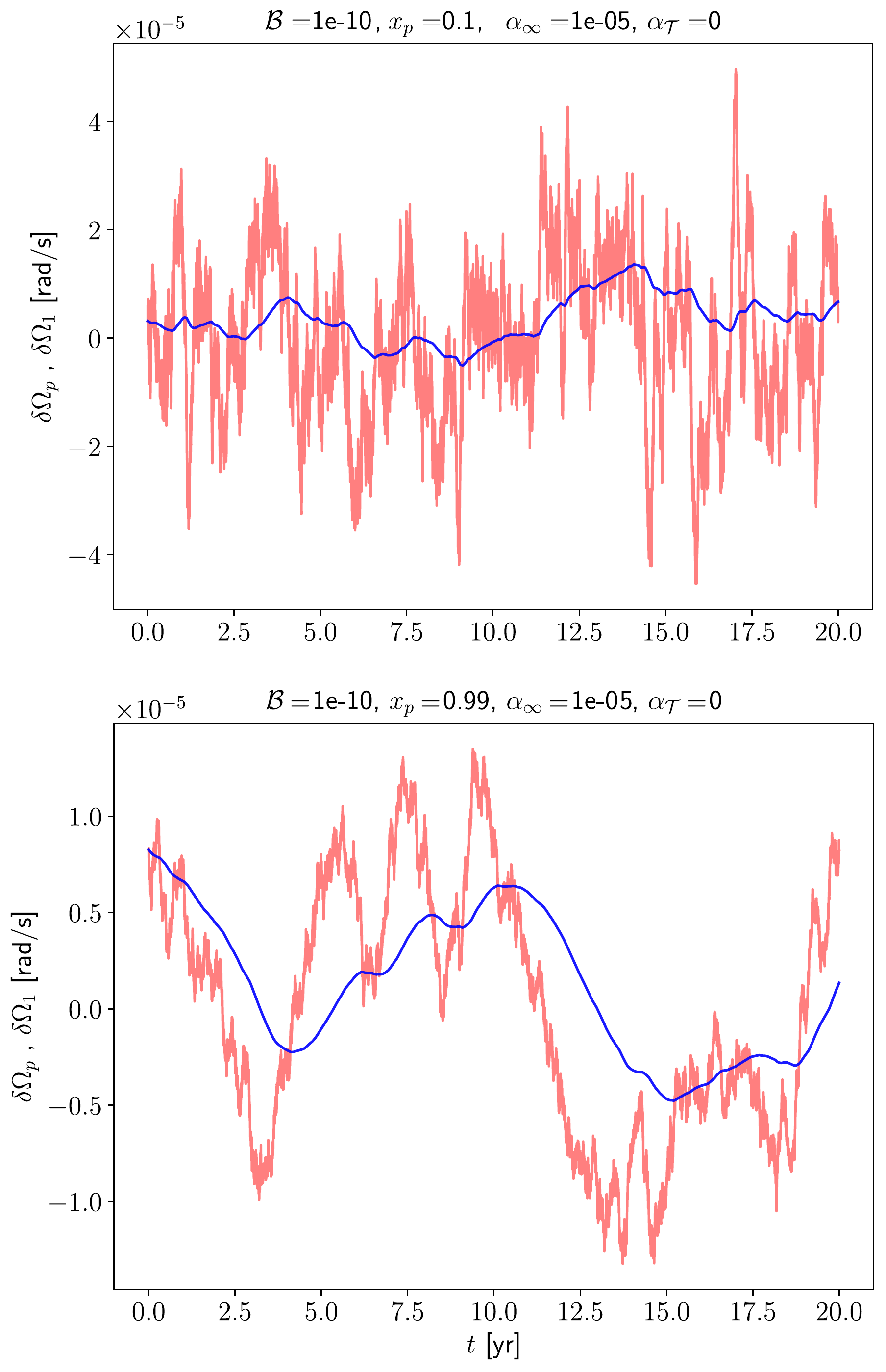}
    \caption{
    A single realisation of the processes $\delta\Omega_p$ (pink) and $\delta\Omega_1$ (dark blue) for the $m=1$ model with fluctuations only in the external torque ($\alpha_\mathcal{T}=0$, $\Omega\approx 70\,$rad/s, $\dot\Omega\approx 10^{-10}\,$rad/s$^2$). The only difference with respect to Fig. \ref{figVela1} is that now  $\mathcal{B}=10^{-10}$. Upper panel: $x_p=0.1$ (the star is mostly superfluid and the resulting relaxation timescale is $\tau \approx 80\,$d). Lower panel: $x_p=0.99$ (the superfluid is confined in the inner crust, $\tau \approx 800\,$d). 
    Since the driving fluctuations are in the external torque, the evolution of the superfluid angular velocity residual $\delta \Omega_1$ lags behind the one of the observable component and is considerably smoother, as expected. }
    \label{figVela2}
\end{figure}

We first consider the case of fluctuations in the external torque ($\alpha_{\mathcal{T}}=0$), see Figs. \ref{figVela1} and \ref{figVela2}. 
In Figure \ref{figVela1} we study the qualitative effect of varying the extension of the superfluid region that participates to the internal torque. In the upper panel we have set $x_p =0.1$, meaning that the star is mostly superfluid: the massive superfluid bulk tends to act as a stabilising reservoir of angular momentum and the resulting $\delta \Omega_p$ process looks almost like a white-noise process. 
On the contrary, when the superfluid reservoir is limited (as in the lower panel of Fig. \ref{figVela1}, $x_p =0.99$), we are qualitatively closer to the $m=0$ case: in this degenerate limit we know that $\delta \Omega_p$ is a Wiener process (i.e. pure Brownian motion, or integrated white-noise). Therefore, the two panels show how changing $x_p$ results in a transition from an evolution that visually looks like a white process to a red one. This is consistent with the fact that the ``white region'' $\omega \in [\mu,\xi]$ where $P_p(\omega)$ is flat reduces to a point for $x_p \approx 1$, see equation \eqref{m2external}.

Another expected feature of the model with pure external noise is that the superfluid component should somehow lag behind the non-superfluid one. In fact, the fluctuations in the external torque act directly on the observable component, that, in turn, exchanges angular momentum with the internal one via the mutual friction. We check this in Fig. \ref{figVela2}, that is completely analogous to Fig. \ref{figVela1} apart from the fact that the dimensionless friction parameter $\mathcal{B}$ is taken to be one order of magnitude smaller. Again, we can visually  recognise that the evolution reddens by increasing the value of $x_p$ but now the long timescale $1/b \approx 800\,$d (common to both the upper and lower panels) makes more evident the fact that the evolution of $\delta \Omega_1$ lags behind $\delta \Omega_p$. This is nothing but the stochastic version of the deterministic steady state solution in \eqref{steady} or \eqref{steady3}, where $\Omega_1$ assumes the same value of $\Omega_p$ exactly after a time $b^{-1}=\tau/x_p$. Since the superfluid lags behind, its evolution is also considerably smoother, as it is the result of a sum of many random kicks imparted by the normal component over the timescale $b^{-1}$.

\begin{figure*}
    \centering
    \includegraphics[scale=0.32]{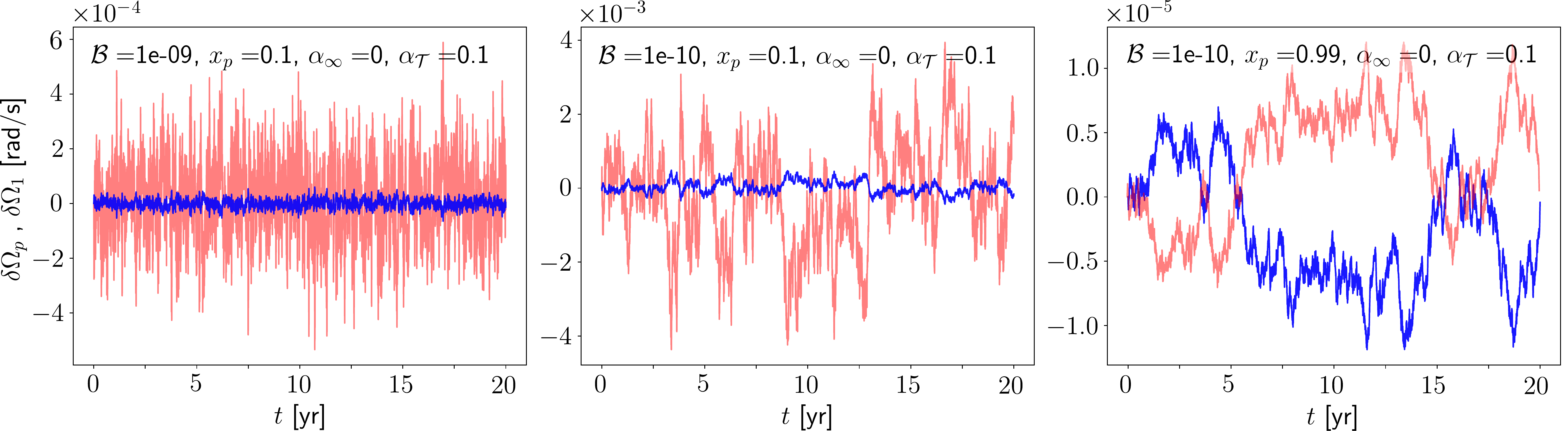}
    \caption{
    A single realisation of the processes $\delta\Omega_p$ (pink) and $\delta\Omega_1$ (blue) for the $m=1$ model with fluctuations only in the internal torque ($\alpha_\infty=0$, $\alpha_\mathcal{T}=0.1$, $\Omega\approx 70\,$rad/s, $\dot\Omega\approx 10^{-10}\,$rad/s$^2$). Right: $\mathcal{B}=10^{-9}$, $x_p=0.1$ ($\tau\approx 8\,$d). Centre: $\mathcal{B}=10^{-10}$, $x_p=0.1$ ($\tau\approx 80\,$d).
    Left: only in this panel the process $\delta\Omega_1$ has been multiplied by $x_1=0.01$ in order to fit it within the canvas ($\mathcal{B}=10^{-10}$, $x_p=0.99$, $\tau\approx 800\,$d), see equation \eqref{pippo}.
    }
    \label{fig_3internal}
\end{figure*}

We briefly comment also on the case with pure internal noise, see Fig. \ref{fig_3internal}, where we have set $\alpha_\mathcal{T}=0.1, \alpha_\infty = 0$. Again, the general trend of reddening in the limit $x_p \rightarrow 1$ still holds true but the evolution of the superfluid is qualitatively different, as it does not lag behind the observable one and it is not smoother. In fact, for pure internal noise the spectral properties of $\delta \Omega_p$ and $\delta \Omega_1$ are identical, since one process is the (rescaled) mirror image of the other, see \eqref{lconst}:
\begin{equation}
\label{pippo}
     \delta{\Omega}_p(t) = - \dfrac{x_1}{x_p} \delta{\Omega}_1 (t) 
    \qquad \quad(\text{i.e.}\,\,\, \delta L =0) \, .
\end{equation}
This is particularly evident in the central and right panels of Fig. \ref{fig_3internal}. 
Let us stress that in all the three figures \ref{figVela1}, \ref{figVela2} and \ref{fig_3internal}, it may look like the lag is reversed when the pink curve is above the blue one. However,  $\delta \Omega_1 -\delta \Omega_p$,  is not the lag, but rather the lag residual with respect to its deterministic value; given our initial condition, the deterministic lag is the constant steady state in \eqref{steady}.

\begin{figure}
    \centering
    \includegraphics[scale=0.11]{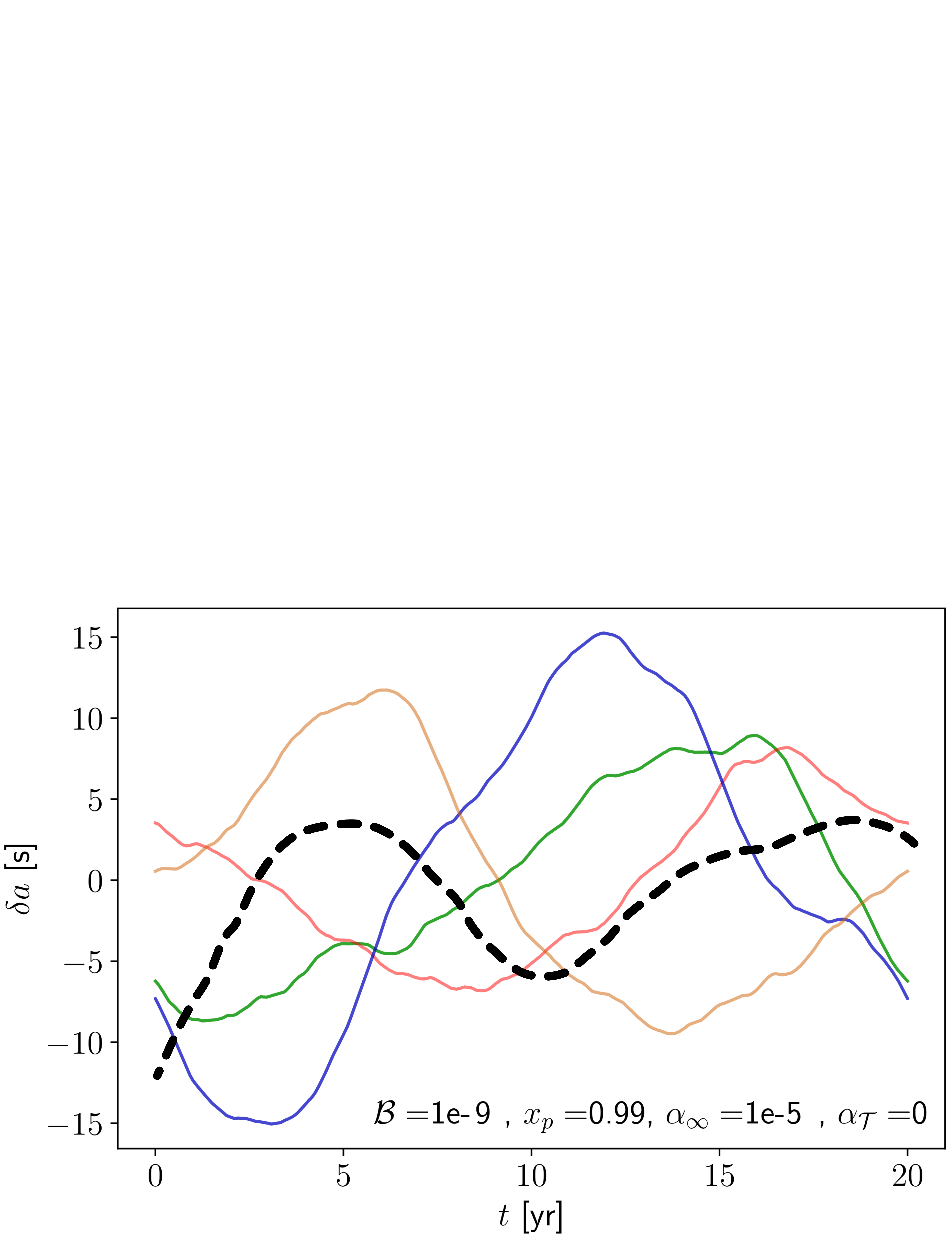}
    \caption{
    Four typical realisations of the timing residuals $\delta a(t)$ for the $m=1$ model with fluctuations only in the external torque ($\alpha_\mathcal{T}=0$, $\Omega\approx 70\,$rad/s, $\dot\Omega\approx 10^{-10}\,$rad/s$^2$, $\mathcal{B}=10^{-9}$). Upper panel: $x_p=0.01$. Lower panel: $x_p=0.99$. The dashed black line represents the timing residuals from 20 years of timing of the Vela pulsar extracted from Fig.~1 of~\citet{shannon_21yr_vela}. }    
    \label{fig_da}
\end{figure}

Finally, it is interesting to to see if the stochastic model can reproduce, at least qualitatively, the 20 years of timing residual data for the Vela pulsar given in Fig. 1  of \citet{shannon_21yr_vela}. Since our model does not include the possible effect of glitches, we have to compare our residuals with the ones reported in panel (b) therein, where the residual arrival times, measured in units of time, are fitted for the glitches. This removes the effect of the glitch recovery from the displayed residuals of \citet{shannon_21yr_vela}; over $20\,$yr the quasi-periodic oscillation of the timing residuals has an amplitude of $\sim 10\,$s. We find that the same features of $\delta a(t)$ can be reproduced by evolving our model with pure external noise, as shown in Fig. \ref{fig_da}. 

{  Following a widespread observational procedure \citep{corders_JPLpulsar_III_1985},  the curves $\delta a(t)$ in Fig. \ref{fig_da} do not start from the origin of the axes because the best-fit quadratic polynomial has been removed. This only modifies their spectrum in the inaccessible infrared region, while their spectral properties for $\omega \gtrsim \nu_o$ are left practically unchanged. From a purely visual point of view, this also eases the comparison of the simulated residuals with the ones of \cite{shannon_21yr_vela}, cf. the Vela residuals in Fig. 2 of \cite{Lower_Utmost_II}. More precise analysis should be done by contrasting the empirical PSD extracted from timing data with the theoretical one, which possibly allows falsifying the assumptions of our theoretical model (e.g., the assumption of uncorrelated white noise terms in the Langevin equation).
} For the moment, we limit ourselves to notice that, for a very low value of $x_p$ (in the upper panel we used the unphysical value $x_p=0.01$), the overall evolution is qualitatively similar, but some small scale decorations (that are absent in the observed timing residuals, the black dotted line in Fig. \ref{fig_da}) are present. This may be a by product of the fact that observations have a lower resolution than our simulations.  

\subsection{Power spectral density}

Even though our simple derivation of the PSD $P_p(\omega)$ is based on the well known trick of computing the Fourier transform of the Langevin equation, this procedure is far from being rigorous in our case \citep{priestley1965} and, more importantly, it is not obvious how well the final $P_p(\omega)$ may approximate the PSD obtained from a finite sample of the signal $\delta \Omega_p$ over a time span $T_o$ by means of the discrete Fourier transform (DFT). 

In fact, the procedure we have used to obtain $P_p(\omega)$ is typically used for mean-reverting multivalued Ornstein-Uhlenbeck processes for which, however, all eigenvalues of $B$ are strictly negative \citep{Gardiner1994book,Singh2018pre}, see also App. \ref{appOU}. 

On the other hand, our matrix $B$ is singular (it has at least one null eigenvector).
On top of this theoretical problem, we also have the well known fact that windowing and sampling of a continuous signal, in this case $\delta \Omega_p$ leads to a distortion of the power spectrum , see e.g. \citep{vanderKlis1989}. Therefore, it is better to check numerically the validity of our result for the PSD. It will be sufficient to do so for the $m=1$ model since, morally, the case $m=2$ is identical: the PSD for both $m=1$ and $m=2$ have the same kind of infrared divergence for $\omega \rightarrow 0$, that ultimately results from long-lived correlations -- the autocovariance is not integrable -- typical of some non-stationary processes, including the Wiener process, see~\citep{Kasdin95}.

\begin{figure}
    \centering
    \includegraphics[scale=0.45]{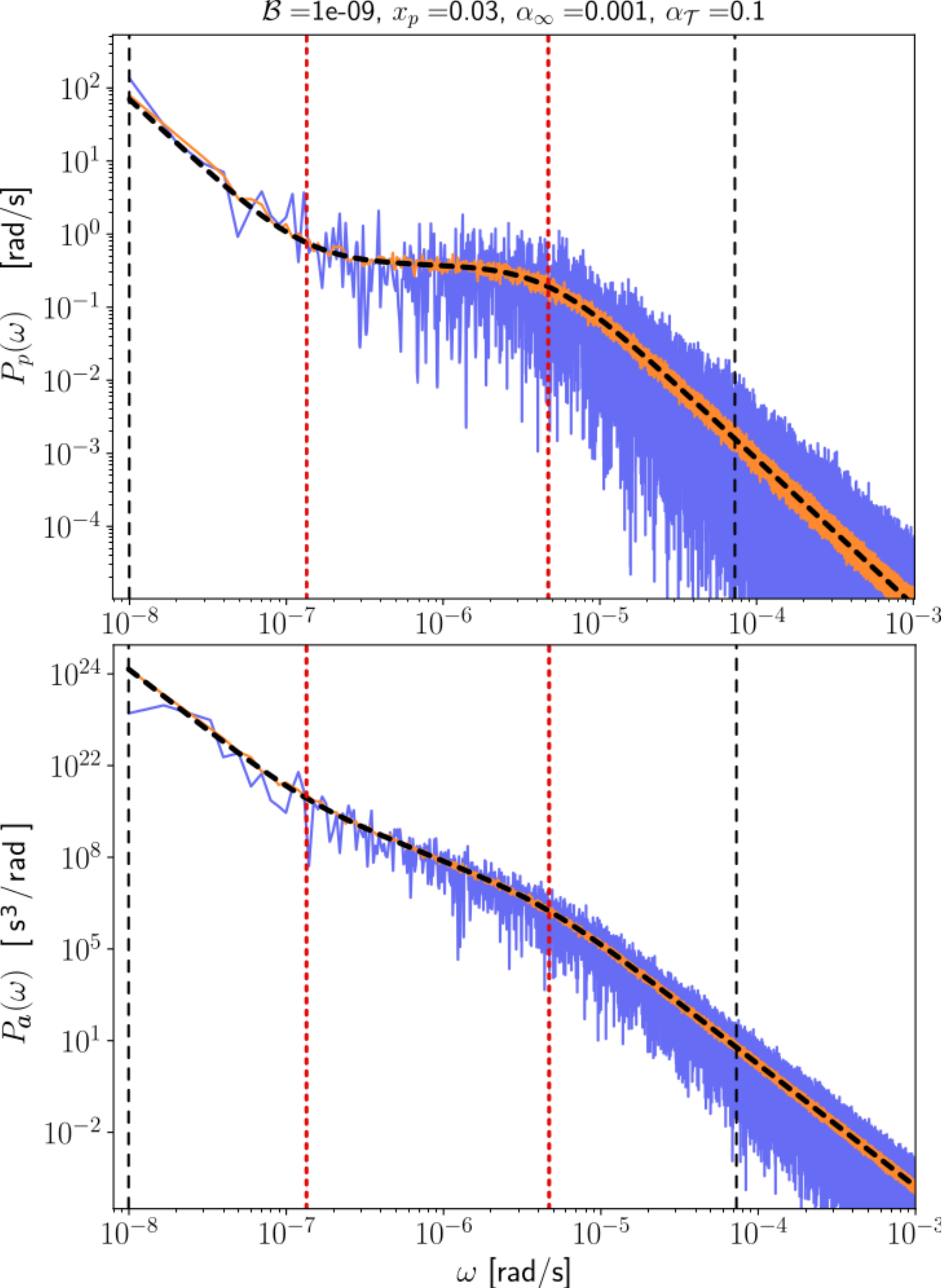}
    \caption{
    Upper panel: comparison between the PSD for the angular velocity  in equation \eqref{P002}, the black dashed line, and the PSD obtained from the discrete Fourier transform of simulated data (orange and blue curves) obtained via equation \eqref{scaling}. The blue curve has been extracted from a single realisation, the orange one obtained after taking the ensemble average of $p_k$ in  \eqref{scaling} over 50 realisations.
    The parameters used are: $\alpha_\infty=10^{-3}$, $\alpha_\mathcal{T}=0.1$, $x_p = 0.03$, $\mathcal{B}=10^{-9}$, $\Omega\approx 70\,$rad/s, $\dot\Omega\approx 10^{-10}\,$rad/s$^2$. Given these parameters, the relaxation time is $\tau\approx 2.5\,$d. 
    The two corner frequencies that define the whitened region $\mu <\omega< \xi$ are indicated by the green dashed lines. The vertical dotted lines correspond to the the first non-DC frequency $\nu_o = 2 \pi/T_o$, for $T_0  = 20\,$ yr, and to a sampling frequency of 1 data point per day, $\nu_s = 2 \pi/$day. Lower panel: the corresponding PSD for the timing residuals $\delta a$.
    }
    \label{fig_psd_vel}
\end{figure}

Given the signal $\delta\Omega_p$, sampled at times $t_l = l T_0/N$ for $l=0,...,N-1$, our exact definition of DFT is,
\begin{equation}
 \delta\tilde{\Omega}_p(\omega_k) =  \, \sum_{l=0}^{N-1}\, 
 \delta{\Omega}_p(t_i) \,  e^{- 2 \pi i \, l k/N}
 \qquad \quad \omega_k = k \nu_s \, ,
\end{equation}
where  the sampling frequency is $\nu_s = 2 \pi N/T_0$ and the first non-DC frequency is 
$\nu_0 = \omega_1 = 2 \pi /T_0$. In this way, the ``negative frequency'' part of the spectrum is contained in the second half of the DFT vector, and we can restrict our attention to $\delta\tilde{\Omega}_p(\omega_k)$ for $k=0,...,N/2$. Clearly, in contrast with he continuum transform $\delta{\Omega}_p(\omega) $, $\delta\tilde{\Omega}_p$ has the same physical dimensions of an angular velocity. Therefore, to obtain something that well approximates $P_p(\omega)$ we have to rescale the discrete power spectrum $p_k=|\delta\tilde{\Omega}_p(\omega_k)|^2$ as
\begin{equation}
\label{scaling}
    P_p(\omega = \omega_k) \approx \dfrac{T_0}{N^2} \, p_k \, .
\end{equation}
 We apply these prescriptions for a benchmark case in Fig. \ref{fig_psd_vel}, where we plot our theoretical PSD \eqref{P002} for the $m=1$ model  together with the one obtained from the DFT of simulated data over a time span $T_0=20$yr (the scaling in \eqref{scaling} is used). The parameters used have been selected to emphasise the presence of the white region (both internal and external fluctuations are switched on). We see that the theoretical result $P_p(\omega)$ approximates well the spectrum obtained by averaging $p_k$ over an ensemble of $50$ realisations of $\delta \Omega_p(t)$. The PSD extracted from a single realisation (that is always the case for real timing data) is considerably more noisy, but the overall shape and scaling is the same. However, the PSD extracted from a single realisation shows that it can be difficult to identify exactly the whitened region in real data, especially if its extent is less than one order of magnitude in $\omega$.
 
 We apply the same strategy to the timing residuals $\delta a$, obtained via the prescription \eqref{semplice}. The corresponding PSD is shown in the lower panel of Figure \ref{fig_psd_vel}: inside the whitened region the spectral index is -2, outside -4.

\section{Conclusions}

We have studied the possibility to describe the timing noise by means of a simple stochastic model for the fluctuations in both the external and internal torques in a neutron star. 
The present framework offers a guideline to interpret the observed features of timing noise in both single pulsars and across pulsar population by including an arbitrary number of internal components.
The framework allows one to model the phenomenology associated to different physical situations: when the source of fluctuations is the mutual friction associated to the motion of superfluid vortices, i.e. fluctuations in the internal torques between the components, and when the source of fluctuations is associated to the external braking torque.

We find that the scaling of the timing noise strength $\sigma \sim \Omega^a \dot\Omega^b$ is different for these two physical scenarios, see Table \ref{tab_sigma}. Moreover, assuming that fluctuations in both the internal and external torques are active, the global scaling of $\sigma$ can not be written just in terms of the two exponents $a$ and $b$, as it is evident in Fig. \ref{figppdot}. In this case, one should consider the full result in \eqref{quellochevoglio} or, at least, the sum of the two leading terms in \eqref{faticafinale}.

Unfortunately, analysis that aim to extrapolate  $a$ and $b$ from observations across the pulsar population is difficult and results reported in the literature agree only at the qualitative level, see Tab.~\ref{tab_sigma}. {  However, taking as a reference the recent estimates of $a$ and $b$ by \citet{Lower_Utmost_II}, we see that our stochastic model could reproduce the timing noise across the pulsar population only if at least one of the three possibilities is met (further work is needed to assess which is the most realistic possibility): injected noise is coloured rather than white (as in the the model for internal turbulence of \citealt{MelatosLink2014}); fluctuations in the internal and external torques are both present; the fluctuation strength varies across the population because it is correlated with the pulsar's age. }
Regarding this last possibility, the physical parameters can indeed evolve as the neutron star cools down (e.g. the phenomenological friction parameter may be higher in younger pulsars because of their higher internal temperature), so that the final scaling $\sigma \propto \Omega^a \dot\Omega^b$ is also affected. However, in the case of pure external torque we have no dependence on the friction parameter in the leading term of $\sigma$, meaning that the overall scaling is less likely to be affected by physical differences due to age.

Similarly, also the power spectra of the angular velocity, phase and TOAs residuals are dependent on the different physical parameter of the neutron star. Therefore, using these analytical results to model the timing noise from the timing data data may help in constraining some global properties of pulsars, like the body averaged friction coupling or the moment of inertia of the components. This provides a way to probe the interior of neutron stars by means of timing noise that is analogous to what is done for pulsar glitches, where observations of the spin up in tandem with a Bayesian analysis based on minimal (deterministic) models of the kind discussed here, allows to extract estimates of the coupling parameters and moments of inertia \citep{montoli2020A&A}. 

We have also tentatively applied our model to estimate the timing noise strength $\sigma$ across the population of known pulsars. Under the working hypothesis that the physical parameters have no dependence the pulsar's age, we have found a general trend where the younger pulsars have much larger timing noise than the older ones (Fig. \ref{figppdot}), in accordance with observations \citep{Arzoumanian1994ApJ,Shannon_noiseMSP_2010}. 
A similar pattern has been observed for the glitch activity, see e.g. \citep{fuentes_2017AA,BSA+2021}. Since the modelling proposed here is closely related to the kind of minimal models used in pulsar glitch studies, it will be interesting to extend the present framework beyond the hypothesis of pure white noise, to include also the effect of glitches. 

Finally, we have reproduced the qualitative behaviour and amplitude of the timing residuals extracted from 20 years of almost phase-coherent monitoring of the Vela pulsar \citep{shannon_21yr_vela}. 
Our limited goal was to start to probe the parameter space of our Langevin model, and we have found that the limit in which the external noise is dominant may be a good starting point for a minimal (in the sense that it contains the least amount of free physical parameters) model for the timing noise in the Vela (see Fig. \ref{fig_da}). 
Of course, such a minimal model will have to be contrasted with real data in a detailed analysis.

\section{Acknowledgements}
Partial support comes from PHAROS, COST Action CA16214. 
M.A. and B.H. acknowledge support from the Polish National Science Centre grant SONATA BIS 2015/18/E/ST9/00577. 
A.B. acknowledges the support from the UK Science and Technology Facilities Council(STFC). 
A consolidated grant from STFC supports the pulsar research at the Jodrell Bank Centre for Astrophysics.

\section{Data Availability}
The numerical results and the code underlying this work are available from~M.A., upon reasonable request. 

\bibliographystyle{mnras}
\bibliography{refer.bib}

\begin{thebibliography}{}
\makeatletter
\relax
\def\mn@urlcharsother{\let\do\@makeother \do\$\do\&\do\#\do\^\do\_\do\%\do\~}
\def\mn@doi{\begingroup\mn@urlcharsother \@ifnextchar [ {\mn@doi@}
  {\mn@doi@[]}}
\def\mn@doi@[#1]#2{\def\@tempa{#1}\ifx\@tempa\@empty \href
  {http://dx.doi.org/#2} {doi:#2}\else \href {http://dx.doi.org/#2} {#1}\fi
  \endgroup}
\def\mn@eprint#1#2{\mn@eprint@#1:#2::\@nil}
\def\mn@eprint@arXiv#1{\href {http://arxiv.org/abs/#1} {{\tt arXiv:#1}}}
\def\mn@eprint@dblp#1{\href {http://dblp.uni-trier.de/rec/bibtex/#1.xml}
  {dblp:#1}}
\def\mn@eprint@#1:#2:#3:#4\@nil{\def\@tempa {#1}\def\@tempb {#2}\def\@tempc
  {#3}\ifx \@tempc \@empty \let \@tempc \@tempb \let \@tempb \@tempa \fi \ifx
  \@tempb \@empty \def\@tempb {arXiv}\fi \@ifundefined
  {mn@eprint@\@tempb}{\@tempb:\@tempc}{\expandafter \expandafter \csname
  mn@eprint@\@tempb\endcsname \expandafter{\@tempc}}}

\bibitem[\protect\citeauthoryear{{Alpar}, {Nandkumar}  \& {Pines}}{{Alpar}
  et~al.}{1986}]{alpar_noise_1986}
{Alpar} M.~A.,  {Nandkumar} R.,   {Pines} D.,  1986, \mn@doi [\apj]
  {10.1086/164765}, \href
  {https://ui.adsabs.harvard.edu/abs/1986ApJ...311..197A} {311, 197}

\bibitem[\protect\citeauthoryear{{Antonelli} \& {Pizzochero}}{{Antonelli} \&
  {Pizzochero}}{2017}]{antonelli2017}
{Antonelli} M.,  {Pizzochero} P.~M.,  2017, \mn@doi [\mnras]
  {10.1093/mnras/stw2376}, \href
  {https://ui.adsabs.harvard.edu/abs/2017MNRAS.464..721A} {464, 721}

\bibitem[\protect\citeauthoryear{{Antonelli}, {Montoli}  \&
  {Pizzochero}}{{Antonelli} et~al.}{2018}]{antonelli2018MNRAS}
{Antonelli} M.,  {Montoli} A.,   {Pizzochero} P.~M.,  2018, \mn@doi [\mnras]
  {10.1093/mnras/sty130}, \href
  {https://ui.adsabs.harvard.edu/abs/2018MNRAS.475.5403A} {475, 5403}

\bibitem[\protect\citeauthoryear{{Arzoumanian}, {Nice}, {Taylor}  \&
  {Thorsett}}{{Arzoumanian} et~al.}{1994}]{Arzoumanian1994ApJ}
{Arzoumanian} Z.,  {Nice} D.~J.,  {Taylor} J.~H.,   {Thorsett} S.~E.,  1994,
  \mn@doi [\apj] {10.1086/173760}, \href
  {https://ui.adsabs.harvard.edu/abs/1994ApJ...422..671A} {422, 671}

\bibitem[\protect\citeauthoryear{{Ashton}, {Lasky}, {Graber}  \&
  {Palfreyman}}{{Ashton} et~al.}{2019}]{ashton2019Nat}
{Ashton} G.,  {Lasky} P.~D.,  {Graber} V.,   {Palfreyman} J.,  2019, \mn@doi
  [Nature Astronomy] {10.1038/s41550-019-0844-6}, \href
  {https://ui.adsabs.harvard.edu/abs/2019NatAs...3.1143A} {3, 1143}

\bibitem[\protect\citeauthoryear{Basu et~al.,}{Basu et~al.}{2021}]{BSA+2021}
Basu A.,  et~al., 2021, \mn@doi [\mnras] {10.1093/mnras/stab3336}, 510, 4049

\bibitem[\protect\citeauthoryear{{Baym}, {Pethick}, {Pines}  \&
  {Ruderman}}{{Baym} et~al.}{1969}]{Baym+1969}
{Baym} G.,  {Pethick} C.,  {Pines} D.,   {Ruderman} M.,  1969, \mn@doi [Nature]
  {10.1038/224872a0}, \href
  {https://ui.adsabs.harvard.edu/abs/1969Natur.224..872B} {224, 872}

\bibitem[\protect\citeauthoryear{{Boynton}, {Groth}, {Hutchinson}, {Nanos},
  {Partridge}  \& {Wilkinson}}{{Boynton} et~al.}{1972}]{boynton_apj_1972}
{Boynton} P.~E.,  {Groth} E.~J.,  {Hutchinson} D.~P.,  {Nanos} G.~P. J.,
  {Partridge} R.~B.,   {Wilkinson} D.~T.,  1972, \mn@doi [\apj]
  {10.1086/151550}, \href
  {https://ui.adsabs.harvard.edu/abs/1972ApJ...175..217B} {175, 217}

\bibitem[\protect\citeauthoryear{{Celora}, {Khomenko}, {Antonelli}  \&
  {Haskell}}{{Celora} et~al.}{2020}]{celora2020MNRAS}
{Celora} T.,  {Khomenko} V.,  {Antonelli} M.,   {Haskell} B.,  2020, \mn@doi
  [\mnras] {10.1093/mnras/staa1930}, \href
  {https://ui.adsabs.harvard.edu/abs/2020MNRAS.496.5564C} {496, 5564}

\bibitem[\protect\citeauthoryear{{Coles}, {Hobbs}, {Champion}, {Manchester}  \&
  {Verbiest}}{{Coles} et~al.}{2011}]{Coles_hobbs_2011MNRAS}
{Coles} W.,  {Hobbs} G.,  {Champion} D.~J.,  {Manchester} R.~N.,   {Verbiest}
  J.~P.~W.,  2011, \mn@doi [\mnras] {10.1111/j.1365-2966.2011.19505.x}, \href
  {https://ui.adsabs.harvard.edu/abs/2011MNRAS.418..561C} {418, 561}

\bibitem[\protect\citeauthoryear{{Cordes} \& {Downs}}{{Cordes} \&
  {Downs}}{1985}]{corders_JPLpulsar_III_1985}
{Cordes} J.~M.,  {Downs} G.~S.,  1985, \mn@doi [\apjs] {10.1086/191076}, \href
  {https://ui.adsabs.harvard.edu/abs/1985ApJS...59..343C} {59, 343}

\bibitem[\protect\citeauthoryear{{Cordes} \& {Helfand}}{{Cordes} \&
  {Helfand}}{1980}]{Cordes1980_III}
{Cordes} J.~M.,  {Helfand} D.~J.,  1980, \mn@doi [\apj] {10.1086/158150}, \href
  {https://ui.adsabs.harvard.edu/abs/1980ApJ...239..640C} {239, 640}

\bibitem[\protect\citeauthoryear{{D'Alessandro}}{{D'Alessandro}}{1996}]{alessandro_review_1996}
{D'Alessandro} F.,  1996, \mn@doi [\apss] {10.1007/BF00637401}, \href
  {https://ui.adsabs.harvard.edu/abs/1996Ap&SS.246...73D} {246, 73}

\bibitem[\protect\citeauthoryear{{Easson}}{{Easson}}{1979}]{easson1979}
{Easson} I.,  1979, \mn@doi [\apj] {10.1086/156842}, \href
  {https://ui.adsabs.harvard.edu/abs/1979ApJ...228..257E} {228, 257}

\bibitem[\protect\citeauthoryear{{Fuentes}, {Espinoza}, {Reisenegger}, {Shaw},
  {Stappers}  \& {Lyne}}{{Fuentes} et~al.}{2017}]{fuentes_2017AA}
{Fuentes} J.~R.,  {Espinoza} C.~M.,  {Reisenegger} A.,  {Shaw} B.,  {Stappers}
  B.~W.,   {Lyne} A.~G.,  2017, \mn@doi [\aap] {10.1051/0004-6361/201731519},
  \href {https://ui.adsabs.harvard.edu/abs/2017A&A...608A.131F} {608, A131}

\bibitem[\protect\citeauthoryear{{Gardiner}}{{Gardiner}}{1994}]{Gardiner1994book}
{Gardiner} C.~W.,  1994, {Handbook of stochastic methods for physics, chemistry
  and the natural sciences}

\bibitem[\protect\citeauthoryear{{Gavassino}, {Antonelli}, {Pizzochero}  \&
  {Haskell}}{{Gavassino} et~al.}{2020}]{gavassino2020MNRAS}
{Gavassino} L.,  {Antonelli} M.,  {Pizzochero} P.~M.,   {Haskell} B.,  2020,
  \mn@doi [\mnras] {10.1093/mnras/staa886}, \href
  {https://ui.adsabs.harvard.edu/abs/2020MNRAS.494.3562G} {494, 3562}

\bibitem[\protect\citeauthoryear{{Graber}, {Cumming}  \& {Andersson}}{{Graber}
  et~al.}{2018}]{Graber2018ApJ}
{Graber} V.,  {Cumming} A.,   {Andersson} N.,  2018, \mn@doi [\apj]
  {10.3847/1538-4357/aad776}, \href
  {https://ui.adsabs.harvard.edu/abs/2018ApJ...865...23G} {865, 23}

\bibitem[\protect\citeauthoryear{{Groth}}{{Groth}}{1975}]{Groth_1975ApJ_III}
{Groth} E.~J.,  1975, \mn@doi [\apjs] {10.1086/190354}, \href
  {https://ui.adsabs.harvard.edu/abs/1975ApJS...29..453G} {29, 453}

\bibitem[\protect\citeauthoryear{{G{\"u}gercino{\v{g}}lu} \&
  {Alpar}}{{G{\"u}gercino{\v{g}}lu} \& {Alpar}}{2017}]{guerci2017MNRAS}
{G{\"u}gercino{\v{g}}lu} E.,  {Alpar} M.~A.,  2017, \mn@doi [\mnras]
  {10.1093/mnras/stx1937}, \href
  {https://ui.adsabs.harvard.edu/abs/2017MNRAS.471.4827G} {471, 4827}

\bibitem[\protect\citeauthoryear{{Haskell}, {Pizzochero}  \&
  {Sidery}}{{Haskell} et~al.}{2012}]{haskell_pizzochero_2012}
{Haskell} B.,  {Pizzochero} P.~M.,   {Sidery} T.,  2012, \mn@doi [\mnras]
  {10.1111/j.1365-2966.2011.20080.x}, \href
  {https://ui.adsabs.harvard.edu/abs/2012MNRAS.420..658H} {420, 658}

\bibitem[\protect\citeauthoryear{{Heintzmann}, {Kundt}  \&
  {Schrufer}}{{Heintzmann} et~al.}{1973}]{Heintzmann_AA_1973}
{Heintzmann} H.,  {Kundt} W.,   {Schrufer} E.,  1973, \aap, \href
  {https://ui.adsabs.harvard.edu/abs/1973A&A....27...45H} {27, 45}

\bibitem[\protect\citeauthoryear{{Hobbs}, {Lyne}  \& {Kramer}}{{Hobbs}
  et~al.}{2010}]{hobbs_2010MNRAS}
{Hobbs} G.,  {Lyne} A.~G.,   {Kramer} M.,  2010, \mn@doi [\mnras]
  {10.1111/j.1365-2966.2009.15938.x}, \href
  {https://ui.adsabs.harvard.edu/abs/2010MNRAS.402.1027H} {402, 1027}

\bibitem[\protect\citeauthoryear{Jankowski, van Straten, Keane, Bailes, Barr,
  Johnston  \& Kerr}{Jankowski et~al.}{2017}]{Jankowski}
Jankowski F.,  van Straten W.,  Keane E.~F.,  Bailes M.,  Barr E.~D.,  Johnston
  S.,   Kerr M.,  2017, \mn@doi [\mnras] {10.1093/mnras/stx2476}, 473, 4436

\bibitem[\protect\citeauthoryear{{Janssen} \& {Stappers}}{{Janssen} \&
  {Stappers}}{2006}]{Janssen_slow30_2006A}
{Janssen} G.~H.,  {Stappers} B.~W.,  2006, \mn@doi [\aap]
  {10.1051/0004-6361:20065267}, \href
  {https://ui.adsabs.harvard.edu/abs/2006A&A...457..611J} {457, 611}

\bibitem[\protect\citeauthoryear{{Johnston} \& {Galloway}}{{Johnston} \&
  {Galloway}}{1999}]{Johnston1999MNRAS}
{Johnston} S.,  {Galloway} D.,  1999, \mn@doi [\mnras]
  {10.1046/j.1365-8711.1999.02737.x}, \href
  {https://ui.adsabs.harvard.edu/abs/1999MNRAS.306L..50J} {306, L50}

\bibitem[\protect\citeauthoryear{{Jones}}{{Jones}}{1990}]{jones_noise_1990MNRAS}
{Jones} P.~B.,  1990, \mnras, \href
  {https://ui.adsabs.harvard.edu/abs/1990MNRAS.246..364J} {246, 364}

\bibitem[\protect\citeauthoryear{Kasdin}{Kasdin}{1995}]{Kasdin95}
Kasdin N.,  1995, \mn@doi [Proceedings of the IEEE] {10.1109/5.381848}, 83, 802

\bibitem[\protect\citeauthoryear{Lam et~al.,}{Lam
  et~al.}{2019}]{Lam_NANOGrav_2019}
Lam M.~T.,  et~al., 2019, \mn@doi [\apj] {10.3847/1538-4357/ab01cd}, 872, 193

\bibitem[\protect\citeauthoryear{{Liu}, {Keith}, {Bassa}  \& {Stappers}}{{Liu}
  et~al.}{2019}]{Liu_2019MNRAS}
{Liu} X.~J.,  {Keith} M.~J.,  {Bassa} C.~G.,   {Stappers} B.~W.,  2019, \mn@doi
  [\mnras] {10.1093/mnras/stz1801}, \href
  {https://ui.adsabs.harvard.edu/abs/2019MNRAS.488.2190L} {488, 2190}

\bibitem[\protect\citeauthoryear{{Lorimer} \& {Kramer}}{{Lorimer} \&
  {Kramer}}{2005}]{Lorimer_book_2004}
{Lorimer} D.~R.,  {Kramer} M.,  2005, Handbook of Pulsar Astronomy.
Cambridge Observing Handbooks for Research Astronomers, Cambridge University
  Press

\bibitem[\protect\citeauthoryear{{Lower} et~al.,}{{Lower}
  et~al.}{2020}]{Lower_Utmost_II}
{Lower} M.~E.,  et~al., 2020, \mn@doi [\mnras] {10.1093/mnras/staa615}, \href
  {https://ui.adsabs.harvard.edu/abs/2020MNRAS.494..228L} {494, 228}

\bibitem[\protect\citeauthoryear{{Lower} et~al.,}{{Lower}
  et~al.}{2021}]{Lower2021mnras}
{Lower} M.~E.,  et~al., 2021, \mn@doi [\mnras] {10.1093/mnras/stab2678}, \href
  {https://ui.adsabs.harvard.edu/abs/2021MNRAS.508.3251L} {508, 3251}

\bibitem[\protect\citeauthoryear{{Lyne}, {Hobbs}, {Kramer}, {Stairs}  \&
  {Stappers}}{{Lyne} et~al.}{2010}]{Lyne+2010}
{Lyne} A.,  {Hobbs} G.,  {Kramer} M.,  {Stairs} I.,   {Stappers} B.,  2010,
  \mn@doi [Science] {10.1126/science.1186683}, \href
  {https://ui.adsabs.harvard.edu/abs/2010Sci...329..408L} {329, 408}

\bibitem[\protect\citeauthoryear{{Manchester}, {Hobbs}, {Teoh}  \&
  {Hobbs}}{{Manchester} et~al.}{2005}]{ATNFcatalogue}
{Manchester} R.~N.,  {Hobbs} G.~B.,  {Teoh} A.,   {Hobbs} M.,  2005, \mn@doi
  [Aj] {10.1086/428488}, \href
  {http://adsabs.harvard.edu/abs/2005AJ....129.1993M} {129, 1993}

\bibitem[\protect\citeauthoryear{{Melatos} \& {Link}}{{Melatos} \&
  {Link}}{2014}]{MelatosLink2014}
{Melatos} A.,  {Link} B.,  2014, \mn@doi [MNRAS] {10.1093/mnras/stt1828}, \href
  {https://ui.adsabs.harvard.edu/abs/2014MNRAS.437...21M} {437, 21}

\bibitem[\protect\citeauthoryear{{Meyers}, {Melatos}  \& {O'Neill}}{{Meyers}
  et~al.}{2021a}]{meyers2021a}
{Meyers} P.~M.,  {Melatos} A.,   {O'Neill} N.~J.,  2021a, \mn@doi [\mnras]
  {10.1093/mnras/stab262}, \href
  {https://ui.adsabs.harvard.edu/abs/2021MNRAS.502.3113M} {502, 3113}

\bibitem[\protect\citeauthoryear{{Meyers}, {O'Neill}, {Melatos}  \&
  {Evans}}{{Meyers} et~al.}{2021b}]{meyers2021b}
{Meyers} P.~M.,  {O'Neill} N.~J.,  {Melatos} A.,   {Evans} R.~J.,  2021b,
  \mn@doi [\mnras] {10.1093/mnras/stab1952}, \href
  {https://ui.adsabs.harvard.edu/abs/2021MNRAS.506.3349M} {506, 3349}

\bibitem[\protect\citeauthoryear{{Montoli}, {Antonelli}, {Magistrelli}  \&
  {Pizzochero}}{{Montoli} et~al.}{2020}]{montoli2020A&A}
{Montoli} A.,  {Antonelli} M.,  {Magistrelli} F.,   {Pizzochero} P.~M.,  2020,
  \mn@doi [\aap] {10.1051/0004-6361/202038340}, \href
  {https://ui.adsabs.harvard.edu/abs/2020A&A...642A.223M} {642, A223}

\bibitem[\protect\citeauthoryear{{Parthasarathy} et~al.,}{{Parthasarathy}
  et~al.}{2019}]{Parthasarathy2020MNRAS_I}
{Parthasarathy} A.,  et~al., 2019, \mn@doi [\mnras] {10.1093/mnras/stz2383},
  \href {https://ui.adsabs.harvard.edu/abs/2019MNRAS.489.3810P} {489, 3810}

\bibitem[\protect\citeauthoryear{{Pizzochero}, {Montoli}  \&
  {Antonelli}}{{Pizzochero} et~al.}{2020}]{Pizzochero2020A&A}
{Pizzochero} P.~M.,  {Montoli} A.,   {Antonelli} M.,  2020, \mn@doi [\aap]
  {10.1051/0004-6361/201937019}, \href
  {https://ui.adsabs.harvard.edu/abs/2020A&A...636A.101P} {636, A101}

\bibitem[\protect\citeauthoryear{Priestley}{Priestley}{1965}]{priestley1965}
Priestley M.~B.,  1965, Journal of the Royal Statistical Society. Series B
  (Methodological), 27, 204

\bibitem[\protect\citeauthoryear{{Rice}}{{Rice}}{1944}]{rice1944}
{Rice} S.~O.,  1944, \mn@doi [Bell System Technical Journal]
  {10.1002/j.1538-7305.1944.tb00874.x}, \href
  {https://ui.adsabs.harvard.edu/abs/1944BSTJ...23..282R} {23, 282}

\bibitem[\protect\citeauthoryear{{Rickett}}{{Rickett}}{1975}]{Rickett1975ApJ}
{Rickett} B.~J.,  1975, \mn@doi [\apj] {10.1086/153501}, \href
  {https://ui.adsabs.harvard.edu/abs/1975ApJ...197..185R} {197, 185}

\bibitem[\protect\citeauthoryear{{Shannon} \& {Cordes}}{{Shannon} \&
  {Cordes}}{2010}]{Shannon_noiseMSP_2010}
{Shannon} R.~M.,  {Cordes} J.~M.,  2010, \mn@doi [\apj]
  {10.1088/0004-637X/725/2/1607}, \href
  {https://ui.adsabs.harvard.edu/abs/2010ApJ...725.1607S} {725, 1607}

\bibitem[\protect\citeauthoryear{{Shannon}, {Lentati}, {Kerr}, {Johnston},
  {Hobbs}  \& {Manchester}}{{Shannon} et~al.}{2016}]{shannon_21yr_vela}
{Shannon} R.~M.,  {Lentati} L.~T.,  {Kerr} M.,  {Johnston} S.,  {Hobbs} G.,
  {Manchester} R.~N.,  2016, \mn@doi [\mnras] {10.1093/mnras/stw842}, \href
  {https://ui.adsabs.harvard.edu/abs/2016MNRAS.459.3104S} {459, 3104}

\bibitem[\protect\citeauthoryear{{Shaw} et~al.,}{{Shaw}
  et~al.}{2022}]{Shaw+2022}
{Shaw} B.,  et~al., 2022, \mn@doi [\mnras] {10.1093/mnras/stac1156}, \href
  {https://ui.adsabs.harvard.edu/abs/2022MNRAS.513.5861S} {513, 5861}

\bibitem[\protect\citeauthoryear{{Singh}, {Ghosh}  \& {Adhikari}}{{Singh}
  et~al.}{2018}]{Singh2018pre}
{Singh} R.,  {Ghosh} D.,   {Adhikari} R.,  2018, \mn@doi [\pre]
  {10.1103/PhysRevE.98.012136}, \href
  {https://ui.adsabs.harvard.edu/abs/2018PhRvE..98a2136S} {98, 012136}

\bibitem[\protect\citeauthoryear{{Sourie} \& {Chamel}}{{Sourie} \&
  {Chamel}}{2020}]{Sourie_vela_2020MNRAS}
{Sourie} A.,  {Chamel} N.,  2020, \mn@doi [\mnras] {10.1093/mnras/staa253},
  \href {https://ui.adsabs.harvard.edu/abs/2020MNRAS.493..382S} {493, 382}

\bibitem[\protect\citeauthoryear{{Sourie}, {Chamel}, {Novak}  \&
  {Oertel}}{{Sourie} et~al.}{2017}]{sourie2017MNRAS}
{Sourie} A.,  {Chamel} N.,  {Novak} J.,   {Oertel} M.,  2017, \mn@doi [\mnras]
  {10.1093/mnras/stw2613}, \href
  {https://ui.adsabs.harvard.edu/abs/2017MNRAS.464.4641S} {464, 4641}

\bibitem[\protect\citeauthoryear{{Taylor}, {Manchester}  \&
  {Huguenin}}{{Taylor} et~al.}{1975}]{Taylor1975ApJ}
{Taylor} J.~H.,  {Manchester} R.~N.,   {Huguenin} G.~R.,  1975, \mn@doi [\apj]
  {10.1086/153351}, \href
  {https://ui.adsabs.harvard.edu/abs/1975ApJ...195..513T} {195, 513}

\bibitem[\protect\citeauthoryear{{van Haasteren} \& {Levin}}{{van Haasteren} \&
  {Levin}}{2013}]{vanHaasteren2013}
{van Haasteren} R.,  {Levin} Y.,  2013, \mn@doi [\mnras]
  {10.1093/mnras/sts097}, \href
  {https://ui.adsabs.harvard.edu/abs/2013MNRAS.428.1147V} {428, 1147}

\bibitem[\protect\citeauthoryear{van~der Klis}{van~der
  Klis}{1989}]{vanderKlis1989}
van~der Klis M.,  1989, Fourier Techniques in X-Ray Timing.
Springer Netherlands, Dordrecht, pp 27--69,
  \mn@doi{10.1007/978-94-009-2273-0_3}

\makeatother
\end{thebibliography}

\appendix

\section{Extension to a generic number of noise terms and spin-down torques}
\label{app1}

The $m=1$ model in \eqref{batuffolino} can be slightly generalised to account for the possibility that part of the total external torque $\mathcal{T}_\infty$ acts directly on the superfluid component,
\begin{equation}
\begin{split}
& x_p \dot{\Omega}_p =  - \mathcal{T} + \mathcal{T}^p_\infty 
\\
&  x_1 \dot{\Omega}_1 = \mathcal{T} + \mathcal{T}^1_\infty\, ,
\end{split}
\label{batuffolino2}
\end{equation}
where $\mathcal{T}_\infty=\mathcal{T}^p_\infty+\mathcal{T}^1_\infty$. This allows to model the possibility of having an extra spin-down torque $\mathcal{T}^1_\infty$ associated with the emission of continuous gravitational waves by oscillation modes in the superfluid component, as proposed in  \citet{meyers2021a,meyers2021b}.

The model in Sec. \ref{theory} can account for $m$ components and (up to) $m+1$ independent additive noise terms ($m$ fluctuating torques due to the friction of each superfluid layer with the unique normal component plus the external torque). However, the matrix form of the equations \eqref{langevin} is more general: if $M$ is a rectangular matrix with $(m+1)$ rows, an arbitrary number of noise terms can be included and the PSD is still the one in \eqref{rice00}. 
This allows to model the situation where part of the spin-down is partially due to gravitational wave emission from the superfluid layers.

Assuming that each of the $m$ superfluid components also have an associated fluctuating external torque, so that the noise terms are up to $(m+1)+m$, it may also be convenient to work with two matrices, $M_\mathrm{e}$ and $M_\mathrm{i}$ that define the external ($\mathrm{e}$) and internal ($\mathrm{i}$) fluctuations, namely
\begin{equation}
\label{langevin_ei}
\begin{split}
    &\dot{\mathbf{\Omega}}_t = B \, \mathbf{\Omega}_t + \mathbf{A}_t + M_\mathrm{e} \dot{\mathbf{w}}^\mathrm{e}_t 
     + M_\mathrm{i} \dot{\mathbf{w}}^\mathrm{i}_t 
    \\
    & \langle \dot{\mathbf{w}}^\mathrm{i}_t \rangle  = \langle \dot{\mathbf{w}}^\mathrm{e}_t \rangle = 0 
    \\
    &\langle \dot{\mathbf{w}}^\mathrm{i}_t  \dot{\mathbf{w}}^\mathrm{i\top}_s \rangle  
    = \langle \dot{\mathbf{w}}^\mathrm{e}_t  \dot{\mathbf{w}}^\mathrm{e\top}_s \rangle  
    =  \delta(t-s) \mathbb{I} 
        \\
    &\langle \dot{\mathbf{w}}^\mathrm{i}_t  \dot{\mathbf{w}}^\mathrm{e\top}_s \rangle  
    = \langle \dot{\mathbf{w}}^\mathrm{e}_t  \dot{\mathbf{w}}^\mathrm{i\top}_s \rangle  
    = 0 
\end{split}
\end{equation}
where $\dot{\mathbf{w}}^\mathrm{i}_t $ and $\dot{\mathbf{w}}^\mathrm{e}_t$ are two independent sets of white noise processes. Using $M_\mathrm{e,i}$ is not strictly necessary, but it allows us to better keep track of the different nature of the torques in the analysis, as we briefly outline.

To guarantee that \eqref{langevin_ei} is consistent with the constraint \eqref{Ldot}, we have to require the analogous of the action-reaction property~\eqref{actionreaction}, 
\begin{equation}
\label{horcrux}
    B^\top\mathbf{x} =0 
    \qquad  \quad
    M_\mathrm{i}^\top\mathbf{x} =0 
    \qquad  \quad
    \text{(action-reaction property)} 
    \end{equation}
as well as the analogous of \eqref{quisquilie}:  in this case the fluctuation $\eta_t^{\infty}$ of the total external torque $\mathcal{T}_\infty$ is given by  
\begin{equation}
    \mathbf{x}^\top \cdot \mathbf{A}_t =  \mathcal{T}_\infty 
    \qquad  
    \mathbf{x}^\top M_\mathrm{e}  \dot{\mathbf{w}}^\mathrm{e}_t = \eta_t^{\infty} 
\end{equation}
From \eqref{langevin_ei} we have that the Fourier components of the angular velocity residuals are given by
\begin{equation}
    \delta\mathbf{\Omega}_\omega = (i \omega \, \mathbb{I}- B)^{-1} 
  ( M_\mathrm{e} \dot{\mathbf{w}}^\mathrm{e}_\omega + M_\mathrm{i} \dot{\mathbf{w}}^\mathrm{i}_\omega )
  \, ,
\end{equation}
implying that the PSD is
\begin{equation}
\label{rice00appendix}
     P_{p}(\omega) \propto \left[ (i \omega\,\mathbb{I} - B)^{-1} 
     (M_\mathrm{i} M_\mathrm{i}^\top + M_\mathrm{e} M_\mathrm{e}^\top) 
     (-i \omega \,\mathbb{I}- B)^{-1} \right]_{pp}  \, .
\end{equation}
It is interesting to note that the spectrum associated with the total angular momentum still has a purely Brownian spectrum also in this more general situation, just its overall power is due to all the external fluctuations: thanks to \eqref{horcrux}, the analogous of \eqref{PSD_LL} is 
\begin{equation}
\label{LLappendix}
     P_{L}(\omega) =
     \frac{\mathbf{x}^\top M_\mathrm{e} M_\mathrm{e}^\top \mathbf{x} }{\omega^2}
\end{equation}
The Euler–Maruyama scheme for the system in \eqref{langevin_ei} is completely analogous to the one in \eqref{em_scheme}:
\begin{equation}
    \begin{split}
  & \delta\mathbf{\Omega}_{t+\Delta t }
    \approx \delta\mathbf{\Omega}_t+ B \,\delta\mathbf{\Omega}_{t } \, \Delta t +  
    \left( M_\mathrm{e} \,\mathbf{X}^\mathrm{e} + M_\mathrm{i} \,\mathbf{X}^\mathrm{i} \right) 
    \sqrt{\Delta t} 
  \\
  &  {X}^\mathrm{e}_j \sim \mathcal{N}(0,1) \qquad \forall j = 0, ..., m
  \\
  & {X}^\mathrm{i}_j \sim \mathcal{N}(0,1) \qquad \forall j = 1, ..., m
  \qquad ({X}^\mathrm{i}_0 =0)\, 
  \end{split}
\end{equation}
that is consistent with the evolution of the total angular momentum 
\begin{equation}
 \delta L_{t+\Delta t }
  = \delta L_t +\mathbf{x}^\top M_\mathrm{e} \,\mathbf{X}^\mathrm{e} \sqrt{\Delta t} 
  + O(\Delta t^{3/2}) \, .
\end{equation}

\subsection{Comparison with \citet{meyers2021a}} 
\label{pizzone}

The $m=1$ model of \citet{meyers2021a} is a particular realisation of the one in \eqref{langevin_ei}, where $M_\mathrm{i}=0$. In fact, this model implements $2$ fluctuating terms, and has all the properties discussed for the system in  \eqref{langevin_ei}, provided that the coupling parameters appearing in $B$ are such that the action-reaction property \eqref{actionreaction} is respected.
Therefore, it is interesting to discuss the general $m=1$ case, which contains both our model of Sec.~\ref{m1} and the one of \citet{meyers2021a} as special cases. 
Let us consider a system defined by
\begin{equation}
\label{2_comp_mattttt}
%
%
\mathbf{\Omega}=\begin{pmatrix}
    \dot{\Omega}_p \\
    \dot{\Omega}_1
    \end{pmatrix}
\quad 
B = \begin{pmatrix}
    -x_1/ \tau & x_1/ \tau \\
    x_p/ \tau & -x_p/ \tau
     \end{pmatrix}
\quad 
\mathbf{A} =     
    \begin{pmatrix}
     -A_p 
     \\
     -A_1 
    \end{pmatrix}
\end{equation}
for some spin-down torques $A_p>0$ and $A_1>0$, namely\footnote{
    The deterministic steady-state is $\dot{\Omega}_p=\dot{\Omega}_1=-|\dot{\Omega}_{\infty}|=-A_p-A_1$, and steady lag is $\Omega_1-\Omega_p=(x_1 A_p+x_p A_1)/( \tau x_1 x_p )\, $.
}
\begin{equation}
\begin{split}
& x_p \dot{\Omega}_p = \frac{x_p x_1}{\tau}(\Omega_1-\Omega_p  )-A_p
\\
& x_1 \dot{\Omega}_1 \, = \frac{x_p x_1}{\tau}(\Omega_p-\Omega_1  )-A_1
\end{split}
\end{equation}
In this simple case additive noise is conveniently introduced via the rectangular matrix, cf.~\eqref{M2},
\begin{equation}
    M=    \begin{pmatrix}
     \sigma_{p}/x_p & 0 &  -\sigma_\mathcal{T}/x_p \\
     0 & \sigma_{1}/x_1 &   \sigma_\mathcal{T}/x_1 \, ,
    \end{pmatrix}
\end{equation}
where $\sigma_{p,1}$ define the strength of the fluctuations associated with the torques $A_{p,1}$.
Consistently with \eqref{LLappendix}, we have that only the external fluctuations define the strength of $P_L$, and it is easy to find that $P_{L}(\omega) = (\sigma_{p}^2+\sigma_{1}^2)/\omega^2$. 
The PSD reads, cf.~\eqref{P002},
\begin{equation}
\label{P000002}
    P_{p}(\omega) = \dfrac{
    (\sigma_\mathcal{T}^2 + \sigma_p^2)\tau^2\omega^2 +x_p^2(\sigma_p^2+\sigma_1^2)
    }{x_p^2 (\tau^2 \omega^4+\omega^2)} \, .
\end{equation}
The corner frequency $\xi = 1/\tau$ is left unchanged by the presence of the extra fluctuating torque, while for $\mu$ we have, cf. \eqref{P002_b},
\begin{equation}
    \mu^2 = 
    \dfrac{x_p^2 \, (\sigma_{p}^2+\sigma_{1}^2) }{\tau^2\,(\sigma_{p}^2 +\sigma_\mathcal{T}^2 )} \, .
\end{equation}
For $\sigma_1=0$ we recover the model in Sec.~\ref{m1} with $\sigma_p = \sigma_\infty$. 
The ordering $0<\mu<\xi$ is maintained either when
\begin{equation}
    \sigma_1^2 \leq \sigma_\mathcal{T}^2 \quad \text{and} \quad 0< x_p<1
\end{equation}
or 
\begin{equation}
\label{zzy}
    \sigma_\mathcal{T}^2  \leq \sigma_1^2 \quad \text{and} \quad x_p^2<\dfrac{\sigma_p^2+\sigma_\mathcal{T}^2}{\sigma_p^2+\sigma_1^2} \, .
\end{equation}
Otherwise, namely for 
\begin{equation}
    \label{zzyk}
    \sigma_\mathcal{T}^2  \leq \sigma_1^2 \quad \text{and} 
    \quad \dfrac{\sigma_p^2+\sigma_\mathcal{T}^2}{\sigma_p^2+\sigma_1^2}<x_p^2<1 \, ,
\end{equation}
the ordering $0<\xi<\mu$ is realised and the PSD behaves as $P_p \sim \omega^{-4}$ in the frequency range $\xi<\omega<\mu$. For pure external noise ($\sigma_\mathcal{T}=0$) we recover the model of \citet{meyers2021a}, see in particular App. A of \citealt{meyers2021b}: both orderings of the corner frequencies are possible depending on the value of $x_p$, as given in \eqref{zzy} and~\eqref{zzyk}.
It is interesting to note that in the case in which the fluctuations in the internal torque are negligible and $\sigma_p \ll \sigma_1$ the only possible ordering is $0<\xi<\mu$, whatever the value of $x_p$.
 
\section{Pure internal noise as an Ornstein-Uhlenbeck process}
\label{appOU}

For pure internal noise (i.e., $\sigma^2_\infty=0$), our model for $\delta \Omega^p_t$ and generic $m>0$ is a multivariate Ornstein-Uhlenbeck process (see, e.g.,  \citet{Singh2018pre} for a contained introduction to the multivariate Ornstein-Uhlenbeck process). 
The Ornstein-Uhlenbeck process is the only nontrivial process that admits a stationary distribution and is both Gaussian and Markovian. Over time, the process tends to drift towards its mean, meaning that the mean acts as an equilibrium for the process, a property often referred to as mean-reverting~\citep{Gardiner1994book,Singh2018pre}.

For $\sigma_\infty \neq 0 $ our process $\delta \Omega_p$ contains a Brownian component that is not mean-reverting and not (wide sense) stationary. Therefore, the analytical estimation of the PSD is a delicate task \citep{priestley1965}: our results are formal and should be checked with the aid of numerical simulations. 
In particular, we can not directly use the Wiener-Khinchin theorem to estimate the PSD \citep{priestley1965}. However, we can analytically check that our formal results are, at least, consistent with the known properties of a mean-reverting Ornstein-Uhlenbeck process when the limit $\sigma_\infty =0$ is taken. 

\subsection{Angular velocity: Ornstein-Uhlenbeck process}

 For simplicity, consider the  $m=1$ case with $\sigma_\infty =0$: by using \eqref{2_comp_mat} and \eqref{M2} into the general solution \eqref{ang_vel_res}, we can obtain the explicit solution for the $p$ component:
\begin{equation}
    \label{2comp_sol}
    \delta \Omega_p(t) = \frac{\sigma_\mathcal{T}}{x_p} \, e^{-t/\tau}  \int_0^t e^{z/\tau } dW_z \, ,
\end{equation}
where $W_z$ is a standard zero-mean one-dimensional Wiener process.
This expression is consistent with the mean-reverting zero-drift Ornstein-Uhlenbeck process defined by the Langevin equation
\begin{equation}
    \label{OU}
    \delta \dot{\Omega}_p(t) = -\dfrac{1}{\tau} \, \delta \Omega_p(t)
    + \frac{\sigma_\mathcal{T}}{x_p} \,   \dot{W}_z \, ,
\end{equation}
where $\dot{W}_z$ is the standard delta-correlated white-noise process. The mean-reversion property is guaranteed by the fact that the relaxation time is positive, $\tau>0$. The mean and covariance are
\begin{equation}
\label{prime}
\begin{split}
     \langle \delta \Omega_p(t) \rangle & = \delta \Omega_p(0) \, e^{-t/\tau} 
    \\
     c(t,u) & =  \langle  [\delta \Omega_p(t)-\langle\delta \Omega_p(t)\rangle ] 
     [\delta\Omega_p(u)-\langle\delta\Omega_p(u)\rangle]
    \rangle 
    \\
      &= \dfrac{\tau \, \sigma_\mathcal{T}^2 }{2 x_p^2 }\left[ e^{-|t-u|/\tau } - e^{-(t+u)/\tau } \right] \, . 
\end{split}
\end{equation}
The first expression is nothing but the mean-reverting property (in the main text we imposed $\delta \Omega_p(0)=0$), while the covariance $c(t,u)$ is obtained from \eqref{isometry}. 
This process admits an asymptotic  stationary state, where the memory of the initial condition is lost. Setting $u= \Delta t +t$, in the double limit $t \gg \tau$ and $t \gg \Delta t$ we have that $c(t,t+\Delta t)$ does not depend on $t$, implying that the process is wide-sense stationary:
\begin{equation}
   \langle \delta \Omega_p(t) \rangle  = 0 
    \qquad \qquad   c(\Delta t) = \dfrac{\tau \, \sigma_\mathcal{T}^2}{2 x_p^2 } \,  e^{-|\Delta t|/\tau } \, . 
\end{equation}
Therefore, the PSD calculated via the Fourier transform of the Langevin equation \eqref{OU} coincides with the one estimated via the Wiener-Khinchin theorem applied to $c(\Delta t)$:
\begin{equation}
\label{psdOU}
    P_p(\omega) = \int_{-\infty}^\infty \! \! d\Delta t \,\, e^{-i \omega \Delta t} \,  c(\Delta t) = \dfrac{\sigma_\mathcal{T}^2 \, \tau^2 }{x_p^2(1+\omega^2\tau^2)} \, ,
\end{equation}
that is exactly the $\sigma_\infty=0$ limit of the PSD in \eqref{P002}. Note that the PSD obtained by Fourier transforming the Langevin equation gives the correct result for the stationary limit of the Ornstein-Uhlenbeck process. Moreover, the average power in the stationary state is 
\begin{equation}
   \int_{-\infty}^\infty \! \dfrac{d\omega}{2 \pi} \,  P_p(\omega)=c(\Delta t=0)=\dfrac{\tau \, \sigma_\mathcal{T}^2}{2 x_p^2 } \, ,
\end{equation}
that coincides with the asymptotic value of the process variance. In other words, we have explicitly checked that the power extracted from the Fourier transform of the Langevin equation coincides with the one obtained from the Wiener-Khinchin theorem in the stationary limit.  
Unfortunately, this does not hold for the integrated process (namely, for the phase), as discussed below.

\subsection{Phase: integrated Ornstein-Uhlenbeck process}
\label{integratedOU}

Since the phase residual $\delta \phi_t$ is just the integrated Ornstein-Uhlenbeck process
\begin{equation}
    \delta \phi_t \, =\, \int_0^t dz \, \Omega_p(z) \, ,
\end{equation}
its associated PSD $P_\phi (\omega)$ is expected to be just \eqref{psdOU} divided by $\omega^2$, see equation \eqref{pphi1}. 
We check to what extent this claim\footnote{ 
    In principle, a more rigorous introduction of the PDS requires the windowing of the signal, so that the Fourier transform of the windowed signal converges~\citep{Kasdin95}.
    } 
about $P_\phi (\omega)$ allows us to estimate the average power of the the signal $\phi_t$, that is directly related to the timing noise strength $\sigma^2$ introduced in \eqref{sigma_continuo}. 
Applying the stochastic Fubini theorem directly to \eqref{prime}, it is possible to show that the average and the covariance are
\begin{equation}
\label{covariancephi}
\begin{split}
     \langle \delta \phi^t \rangle & = \int_0^t \!\!dz\,\langle \delta\Omega_p(z)\rangle =\tau \, \delta\Omega_p(0) \, \left( 1-e^{-t/\tau} \right) 
    \\
     C(t,u) &=  \langle  [\delta \phi_t -\langle\delta \phi_t \rangle ] 
     [\delta \phi_u -\langle\delta \phi_u \rangle]
    \rangle 
    = \dfrac{\tau^2 \, \sigma_\mathcal{T}^2 }{ x_p^2 }\,  \min(t,u) \, +
    \\& + \dfrac{\tau^3 \, \sigma_\mathcal{T}^2 }{2 x_p^2 }
    \left[ 2 e^{-t/\tau} +  2e^{-u/\tau} -2 -e^{-|t-u|/\tau } -\tau e^{-(t+u)/\tau }\right]
\end{split}
\end{equation}
Now, the above expression of the covariance $C$ tells us that $ \delta \phi_t $ has no stationary limit. 
For example, the variance depends on $t$ even in the limit $t\gg \tau$, 
\begin{equation}
\label{variance_integrated_asym}
     C(t,t) \approx  \dfrac{\tau^2 \, \sigma_\mathcal{T}^2 }{  x_p^2 }\, \left( t - 6 \tau \right) + O\left( e^{-t/\tau}\right) .
\end{equation}
However, for the sake of extracting the timing noise strength $\sigma^2$, we can take the average power on a long time interval of duration $T_o$ see equation \eqref{sigma_continuo}: 
we write $\sigma^2$ in terms of the process's variance 
\begin{equation}
    \sigma^2 =\frac{1}{T_o \,  \Omega^2} \int_0^{T_o}\!\! dt\, C(t,t) 
    \approx  \dfrac{T_0 \, \tau^2 \, \sigma_\mathcal{T}^2 }{2 \, x_p^2 \Omega^2 } 
- \dfrac{3 \,\tau^2 \, \sigma_\mathcal{T}^2\, \dot{\Omega}^2\,x_p}{2 \, x_p^2 \Omega^2 }
    \, ,
\end{equation}
where we have taken the limit $T_0 \gg \tau$. This limit is equivalent to $\nu_0 \ll \mathcal{B} \Omega$, where $\nu_0 \sim 1/T_0$ is the infrared cut-off in \eqref{quellochevoglio}. 
If we use the noise parametrization in \eqref{psico}, we finally obtain  
\begin{equation}
    \sigma^2     \approx  
    \dfrac{T_0 \,x_1^2 \, \alpha_\mathcal{T}^2\, \dot{\Omega}^2  }
    {16 \, \mathcal{B}^3 \Omega^5 } \, 
    \left( \, 1- \dfrac{3  \,x_p}{2 \, T_0\, \mathcal{B}\,\Omega}  \, \right)
\end{equation}
Therefore, apart from constant (dimensionless) numerical factors, the timing noise strength associated to the variance of the timing residual is expected to scale as, cf. with \eqref{phi2internal}, 
\begin{equation}
    \sigma^2    \propto   
    T_0  \, \dot{\Omega}^2 \, \Omega^{-5 } \,  
    \propto \, \nu_0  \, \dot{\Omega}^2 \, \Omega^{-5 } \, 
    \qquad \quad (\text{for} \quad \nu_0 \ll \mathcal{B}\Omega )
\end{equation}
This is consistent with the scaling in \eqref{faticafinale}, see also Table \ref{tab_sigma}: 
we have explicitly checked that the band-limited integration of the PSD in equation \eqref{quellochevoglio} leads to an estimation of the timing noise strength that has the same scaling properties of the one estimated directly from the averaged variance of the signal. This is obvious when the PSD is integrable over the whole real line, namely when no infrared cutoff is needed $\nu_0=0$, as in equation \eqref{psdOU}. 


\bsp	
\label{lastpage}
\end{document}